\documentclass[manuscript,nonacm]{acmart}
\usepackage{multirow}

\newcommand{\red}{\textcolor{black}}

\AtBeginDocument{%
  \providecommand\BibTeX{{%
    \normalfont B\kern-0.5em{\scshape i\kern-0.25em b}\kern-0.8em\TeX}}}

\setcopyright{none}
\acmDOI{10.1145/1122445.1122456}

\acmConference[]{}{}{}
\acmBooktitle{}

\begin{document}

\title{Survey of Control-Flow Integrity Techniques for Embedded and Real-Time Embedded Systems}

\author{Tanmaya Mishra}
\email{tanmayam@vt.edu}
\affiliation{%
  \institution{Virginia Polytechnic Institute and State University}
  \country{USA}
  \postcode{22311}
}

\author{Thidapat Chantem}
\email{tchantem@vt.edu}
\affiliation{%
  \institution{Virginia Polytechnic Institute and State University}
  \country{USA}
  \postcode{22311}
}
\author{Ryan Gerdes}
\email{rgerdes@vt.edu}
\affiliation{%
  \institution{Virginia Polytechnic Institute and State University}
  \country{USA}
  \postcode{22311}
}

\renewcommand{\shortauthors}{Mishra, et al.}

\begin{abstract}
  Computing systems, including real-time embedded systems, are becoming increasingly connected to allow for more advanced and safer operation. Such embedded systems are resource-constrained, such as lower processing capabilities, as compared to general purpose computing systems like desktops or servers. However, allowing external interfaces to such embedded systems increases their exposure to attackers. With an increase in attacks against embedded systems ranging from home appliances to industrial control systems operating critical equipment that have hard real-time requirements, it is imperative that defense mechanisms be created that explicitly consider such resource and real-time constraints constraints. Control-flow integrity (CFI) is a family of defense mechanisms that prevent attackers from modifying the flow of execution. We survey CFI techniques, ranging from the basic to state-of-the-art, that are built for embedded systems and real-time embedded systems and find that there is a dearth, especially for real-time embedded systems, of CFI mechanisms. We then present open challenges to the community to help drive research in this domain. 
\end{abstract}




\maketitle

\section{Introduction}
\label{sec:introduction}
Today, computing systems communicate through a complex web of interconnections. For instance, the modern smartphone can simultaneously capture photographs and videos at quality rivaling that of movie cameras, upload gigabytes of information to the internet, turn on lamps and automatically control thermostats, stream high fidelity music to the nearest speaker, and can even unlock a car. We now live in the age of internet-of-things (IoT~\cite{atzori2010internet}) where the physical world around us can be manipulated by a push of a button. 

The convenience afforded by such interconnections is, unfortunately, countered by the inconvenience of dealing with malicious parties who try to take control of these connected devices to inflict monetarily, and in some cases, bodily harm. A simple smart bulb from a reputed company was exploited to launch a distributed denial-of-service (DDOS) attack~\cite{philips}. While a DDOS attack may have, at the most, an economic impact on the victim, an attacker could reprogram the lights to blink so as to induce an epileptic attack in some individuals. Unfortunately, such instances of malicious behavior are not confined to small home appliances. Stuxnet~\cite{falliere2011w32} is a computer worm built to infect supervisory control and data acquisition (SCADA) systems. Infections of this worm were first uncovered in 2010, by when it had already infected nuclear reactor control systems and caused significant damage to Iran's nuclear program. Malicious entities could, theoretically, cause the reactors to fail and cause catastrophic damage to both life and property. Interestingly, many Stuxnet systems were air-gapped, i.e, did not have a direct connection to external systems. Instead, the infection spread from physical drives inserted by human operators. 

However, the scope of system attacks and their related defenses has an extremely large body of prior work. Over the years, a variety of system defense mechanisms have been proposed for a wide range of threat models and system configurations. These mechanisms can be either hardware assisted~\cite{frassetto2017jitguard}, entirely in software~\cite{wang2010taintscope}, implemented in the system pre-deployment, such as compiler-based protections~\cite{10.1145/3141234}, detect attacks during system runtime~\cite{mahsasimple}. \red{Discussing the entire body of work of such defenses is beyond the scope of this survey. Here, we discuss a specific set of defense mechanisms called control-flow integrity (CFI) for embedded, and particularly, real-time embedded systems.  
}
Our major contributions are:

\begin{enumerate}
    \item We explore a number of recently proposed mechanisms targeting embedded systems, specifically those that are \textit{resource-constrained}, such as reduced processing capabilities over general-purpose processing environments systems such as those found in desktop or server-grade equipment. Such embedded systems usually feature low-end processing environments such as microcontrollers (and their related underlying processor architecture) and identify key techniques that could provide inspiration for more robust real-time systems CFI design.
    \item We find that there are very few CFI mechanisms built specifically for real-time embedded systems. Our exploration of the work for embedded systems show that there is an avenue to extend CFI techniques from general embedded systems to create powerful CFI mechanisms that uniquely leverage real-time requirements. 
    \item We consolidate our findings and present challenges and suggestions for future research in Section~\ref{sec:summary_challenges}.
\end{enumerate}

We give definitions for \textit{embedded systems} and \textit{real-time embedded systems} later in this section that defines the scope of our survey. We will now provide an overview of the type of attacks that are countered by CFI and an overview of CFI itself. 

\subsection{\red{Scope of Attacks and Defenses: Control-Flow Attacks and CFI}}
\red{To aid the discussion of CFI, which is the main focus of this survey, it is necessary to first describe the type and scope of attacks for which they are built. This family of attacks are collectively called control-flow attacks. We shall now discuss these types of attacks.
}

\red{\subsubsection{Control-Flow Attacks:} Control-flow attacks capture and modify the \textit{flow of execution} of a program. These attacks attack \textit{control information}, that is information presented to a program during runtime that determines the path that a program takes to continue execution. A simple example of such information is the return address of a function call. See Figure~\ref{fig:canariesshadow}.a for an example of the stack frame of a function call on a generic ARM architecture-based microcontroller. Here, the value stored in the LR field of the stack frame is popped into the special \texttt{LR} or link register. ARM calling convention~\cite{earnshaw2005arm}, which is implemented by all compilers that officially support this architecture, utilizes the \texttt{LR} to implement the return sequence of a function call. Return sequences are implemented by branching on the \texttt{LR} such as by using the \texttt{BX LR} instruction. Therefore, the contents of the \texttt{LR} effectively constitutes control information. Control-flow attacks aim to modify such information to redirect program execution for malicious purposes. The same figure showcases a sample attack where the attacker utilizes a buffer-overflow bug in the code that writes to a memory buffer in the stack frame, such that it uses the bug to overwrite the \texttt{LR} information thereby tainting the return address with a desired target address. Therefore, when the function returns, the tainted value is popped and becomes the target of the branch statement. Note that since control-flow attacks redirect program execution, they are also sometimes called \textit{code redirection} attacks.}

\red{Two broad categories exist in control-flow attacks. While each category is a large research domain by itself, we briefly describe them here for context:
\begin{enumerate}
  \item Code injection attacks - The sample attack we discuss above is a simple example of a code injection attack. As discussed above, the attack can be broken into two stages that are a) injecting (writing) code into some form of executable memory, followed by b) a redirection to the beginning of the injected code, such as by using the \texttt{LR} register. Due to code injection that takes place in stage a), these attacks are termed code injection attacks. Code injection attacks have a large body of work~\cite{ray2012defining,francillon2008code,giannetsos2010arbitrary}. However, such attacks have lost favor over time with advancements in software and hardware architecture. Note that an implicit assumption of the attack is that code is injected into executable memory, that is, the stack is executable. Therefore, to defeat such attacks, it is sufficient to introduce countermeasures that ensure that writeable memory addresses are not executable. A large body of research has been presented to counter code injection attacks with relatively inexpensive performance overheads~\cite{lee2004enlisting,hu2006secure}. Even for lower-end processors, such as microcontrollers from the ARM Cortex-M family, prior work have implemented defenses~\cite{kwon2019uxom}. Modern hardware now include architectural features such as the memory protection unit (MPU) that make it trivial for system designers to implement writeable but non-executable memory, an important requirement for code injection attacks to propagate~\cite{clements2017protecting}. Since such attacks can be defended against relatively easily, such attacks are outside the scope of this survey. 
  \item Code reuse attacks - With the addition of defenses against a code injection attacks, a new class of attacks emerged that are collectively called code reuse attacks. These attacks are a logical extension of code injection attacks where attackers modify control information to reuse arbitrary sequences of code already present within the program binary to perform malicious operations. One of the most famous examples of code reuse attacks is the the return-oriented programming or ROP~\cite{shacham2007geometry, roemer2012return, weidler2019return} attack. We provide more details of ROP, and defenses against such attacks, in Section~\ref{sec:basic_techniques}. Increasingly sophisticated variants of ROP~\cite{carlini2014rop}, such as some that do not even require return sequences~\cite{checkoway2010return, davi2010return} have been proposed over the past decade. A large set of defenses have also been proposed to counter ROP and related attacks~\cite{cheng2014ropecker, volckaert2015cloning, huang2012dynamic}, showcasing the relevance and danger such type of attacks represent for modern systems. 
\end{enumerate}
} 

Since control-flow attacks modify the control-flow of the program, it is necessary to maintain the \textit{integrity} of the control-flow by detecting malicious control-flow deviation when it occurs. Therefore, any defense mechanism that enforces this integrity is called control-flow integrity (CFI)~\cite{abadi2009control}. In this survey, we discuss CFI that is specifically designed to defend against code reuse attacks. Note that we alternatively refer to CFI as CFI enforcement, CFI mechanism, or CFI technique throughout this paper.

\subsubsection{Control-Flow Integrity (CFI):} CFI is the set of system security techniques built to prevent an attacker from forcing a software system to execute code in an unintended manner. CFI focuses on ensuring that system code does not deviate from known software control-paths during system runtime. CFI mechanisms are built to address powerful threat models where it is assumed that the attacker can bypass all other defenses to infiltrate the system and force system software to execute in an arbitrary manner. There is a wealth of research in recent years that develop CFI mechanisms for increasingly complex and powerful attack scenarios~\cite{burow2017control, de2017survey}. CFI mechanisms are also available in many commercial and production-grade software. For example, the Clang compiler implements control-flow violation detection mechanisms~\cite{clang}, and Microsoft has its own CFI implementation called Control Flow Guard~\cite{lastnameholiu} for its Windows operating system which has been available since Windows 8.1. 

While the literature concerning CFI mechanisms (and techniques to bypass them~\cite{ElSarraf2013UsabilityOE,controlflowbend}) is rich with studies regarding the non-negligible performance and/or memory overhead of the mechanisms, few are built specifically for embedded systems and even fewer explicitly consider the real-time requirements of such systems. Therefore, we shall first look at CFI mechanisms for general embedded systems and then move towards mechanisms built explicitly for those with real-time constraints. We shall look at both software-based and hardware-assisted mechanisms, as well as a mechanism that takes advantage of the predictability of real-time systems. However, before we begin discussion of CFI techniques, we shall now define resource and real-time constraints. 

\subsection{Systems Considered: Embedded and Real-Time Embedded Systems}
\label{sec:resource_real_time_constraints}

\red{There exists numerous prior work that are excellent surveys and compilations regarding CFI defenses for general systems~\cite{de2017survey,burow2017control,abadi2009control,Tang2017EternalWI,Sayeed2019ControlFlowIA}. However none of these work explicitly consider system capabilities and constraints.}. We now define the types of systems that we consider for the rest of this work, and their constraints that influence the design of CFI for such systems.

\subsubsection{Embedded Systems:} As discussed earlier, the Stuxnet worm was built specifically to target and control SCADA systems. A SCADA system is usually composed of a number of embedded computing systems built for specific operations, such as data gathering and actuator control. However, embedded computing systems themselves can be found in a wide variety of operating environments, ranging from complex SCADA systems, to robots used for medical procedures as well as small household appliances. These embedded systems are usually severely \textit{resource-constrained} to minimize size, weight and power (SWaP), cost and/or simplify operations. Typically, they consist of microcontrollers that are low-end processors with integrated memory, executing software built to perform specific operations in a deterministic and predictable manner. For example, the modern vehicle can have over a hundred individual computing units, called Electronic Control Units (ECU) that control different functionalities of the vehicle. These units usually consist of a microcontrollers~\cite{krishnadas2016concept} that operate at a clock frequency an order of magnitude lower than the processors found in modern internet servers, and have similarly small amounts memory for storage and operation. These computing units control vehicle operations ranging from non-critical infotainment systems, to extremely critical Advanced Driver Assistance Systems (ADAS), such as anti-lock braking systems, whose failure could result in passenger loss-of-life. Further, the software for such systems may not be regularly updated due to the inaccessibility of their deployment locations. Therefore, once security vulnerabilities in the software are found, they may not be easily patched, making them lucrative targets for malicious entities. In addition, in the case of modern vehicles, increasing inter-vehicular connectivity to improve ADAS as well as increasing number of interfaces such as WiFi and Bluetooth for passenger convenience, has widened the attack surface that can be exploited by such entities~\cite{checkoway2011comprehensive}. Therefore, due to the wide range of applications of resource-constrained embedded systems and their increasing attack surface due to system inter-connectivity, it is imperative that such systems have built-in defense mechanisms to prevent their exploitation by attackers. 

\red{To summarize, our definition of embedded system is in the broader sense. That is, our definition encompasses embedded systems with fixed system resources (memory, processor, peripherals, etc.) where processing elements are embedded off-the-shelf microcontroller architectures such as ARM Cortex-M and ARM Cortex-R~\cite{zlatanov2016arm} or bespoke architectures that evolve from those that could be utilized in similar systems. Such processing environments have fewer architectural features than desktop or server grade processors and usually paired with slower/limited memory and peripherals for managing costs and/or special memory systems for redundancy and safety. Such systems are usually deployed in mission-specific applications in a wide range of domains, such as industrial, automotive, space and medical systems or even internet-of-things (IoT) systems. Our definition of such systems is broad since it allows us some flexibility to look at CFI mechanisms that may work for a specific type of embedded system, but could be applied to similar architectures with some modifications, giving us a broader field-of-view of the domain. For each mechanism we take a closer look at in later sections, we state the specific architectural considerations that informed its design. Note that for completeness we also briefly discuss some techniques that use external processing resources such in Section~\ref{sec:off_chip} and show their fundamental similarities with techniques that do not require external processing resources. However, we do not present in-depth information for these techniques since they utilize external processing resources that makes it difficult to compare with techniques that do not require such external resources. Note that our definition of embedded systems assumes that such systems are resource-constrained and we interchangeably refer to embedded systems as \textit{embedded systems} or \textit{resource-constrained embedded systems} throughout this work.}

\subsubsection{Real-time embedded systems: } Similarly, many such resource-constrained embedded systems require real-time guarantees. In the case of ADAS systems such as anti-lock braking systems, for example, multiple control loops (including actuator control) must be completed per second to maintain safe vehicle operation. We term such embedded systems as \textit{real-time embedded systems}. If such a system misses any deadline, regardless of the correctness of the computation, the consequences could include the loss of life. When such guarantees are required atop resource-constraints, developing defense mechanisms for such systems become especially challenging.

\red{We therefore focus on defense mechanisms that are built for embedded systems and real-time embedded systems. Since such systems have both \textit{resource} and \textit{real-time} constraints, considering systems that have a combination of these two types of constraints leads to a unique set of problems for designing useful CFI mechanisms for such systems. In general, some of the problems are: 
\begin{enumerate}
  \item Weaker processing capabilities as compared to general-purpose desktop or server grade systems constrains the complexity of the design and scope of the CFI mechanism that can be introduced in the system. Complex CFI would introduce unmanageable overheads that would break the real-time guarantees of the system. For example most of the defenses we discuss specifically for embedded systems in Section~\ref{sec:cfi_resource_constrained} detect irregularities in branch source and targets, individually for each branch. However, general-purpose architectures have more complex mechanisms available~\cite{cheng2014ropecker, Crane2015ReadactorPC} since such systems are not constrained by real-time guarantees and can accept greater performance reductions for higher degree of security. 
  \item In addition, due to reduced hardware capabilities, certain defense mechanisms that are built for general-purpose systems may not be directly applicable to resource-constrained embedded systems. For example some defenses~\cite{cheng2014ropecker} require advanced memory management features such as virtual memory which are not available on low-end microcontrollers. Therefore defenses for such systems require hardware/software workarounds to maintain acceptable levels of defense without hampering real-time operation. 
  \item Real-time systems require a study of the increased overhead due to CFI mechanisms, and its impact on the schedulability of the system. Many CFI techniques designed for general-purpose and, in fact, as we see later in Section~\ref{sec:cfi_resource_constrained}, resource-constrained embedded systems do not discuss schedulability, nor do they discuss possible security-schedulability trade-offs that may be required to balance timing and security. 
  \item Other system parameters, such as power consumption are rarely considered when discussing CFI. Many resource-constrained embedded systems may have access to limited (such as battery-based) or intermittent (such as via renewable energy like solar) power supply. Such constraints are rarely discussed by prior work. We, in fact, realize a gap in knowledge with respect to impact of CFI and power consumption and suggest readers to explore this domain in future work (see Section~\ref{sec:summary_challenges}).
\end{enumerate}
}

To the best of our knowledge, this survey is the first to identify a gap in research of CFI mechanisms for real-time embedded systems, and propose future research avenues that could be considered by the real-time systems community. \red{In this survey, we discuss CFI for real-time embedded systems and not general real-time systems that do not consider resource-constraints (such as memory or low end computation environments) typical to embedded systems. This is due to a lack of CFI literature that explicitly considers real-time constraints \textit{without} considering resource constraints. On the other hand, we believe that our discussion of CFI for real-time embedded systems provides adequate coverage of possible techniques that can be utilized, without many modifications, for any general real-time system. We also believe there is ample opportunity to investigate the unique hardware-software constraints of resource-constrained embedded systems and utilize real-time execution characteristics to aid the development of CFI techniques which are equally applicable to real-time systems that do not suffer from resource-constraints. Such timing based co-design, as we show in later sections, is severely lacking and we present a few possible paths of investigation for the reader to follow for future work in Section~\ref{sec:summary_challenges}. }

\red{\section{Paper Organization}}
\red{The rest of this work is divided into 4 major sections. These are:
\begin{enumerate}
    \item CFI Techniques for Backward and Forward-Edges (Section~\ref{sec:basic_techniques}) - We discuss different CFI designs, from both theoretical and practical approach, for general-purpose systems. This section provides the reader a general overview of how state-of-the art CFI mechanisms, both basic and advanced, are usually designed and implemented.
    \item CFI for Embedded Systems (Section~\ref{sec:cfi_resource_constrained}) - We discuss different types of CFI techniques built specifically for resource-constrained embedded systems. Please note our definition of embedded systems is provided in Section~\ref{sec:resource_real_time_constraints}. As stated in Section~\ref{sec:resource_real_time_constraints}, the nomenclature ``embedded systems" and ``resource-constrained embedded systems" are synonymous and interchangeably used depending on context for clarity. 
    \item CFI for Real-Time Embedded Systems (Section~\ref{sec:cfi_real-time}) - We then discuss how real-time considerations play into the design of CFI for embedded systems. Four specific techniques are considered that explicitly consider real-time constraints and discuss schedulability-security trade-offs and/or schedulability analyses. 
    \item Summary and Open Challenges (Section~\ref{sec:summary_challenges}) - We summarize our discussion of different CFI techniques and discuss some challenges from a real-time perspective, and from overall resource-constrained embedded system perspective.
\end{enumerate}
Table~\ref{tab:table_of_contents} provides a brief overview of the relevant sections where we discuss specific CFI techniques, especially for Section~\ref{sec:forwardedge}, Section~\ref{sec:cfi_resource_constrained} and Section~\ref{sec:cfi_real-time}.
}
\section{CFI techniques for backward and forward-edges}
\label{sec:basic_techniques}

\begin{table*}[]
\begin{tabular}{|l|l|l|l|l|}
\hline
\multicolumn{1}{|c|}{\multirow{2}{*}{\begin{tabular}[c]{@{}c@{}}CFI\\ Mechanisms\end{tabular}}} & \multicolumn{2}{c|}{Forward-edge}                                                                                                                             & \multicolumn{1}{c|}{\multirow{2}{*}{\begin{tabular}[c]{@{}c@{}}Backward-\\ edge\end{tabular}}} & \multicolumn{1}{c|}{\multirow{2}{*}{Mechanism highlights}}                                                            \\ \cline{2-3}
\multicolumn{1}{|c|}{}                                                                          & \multicolumn{1}{c|}{\begin{tabular}[c]{@{}c@{}}Fine-\\ grained\end{tabular}} & \multicolumn{1}{c|}{\begin{tabular}[c]{@{}c@{}}Coarse-\\ grained\end{tabular}} & \multicolumn{1}{c|}{}                                                                          & \multicolumn{1}{c|}{}                                                                                                 \\ \hline
\multicolumn{5}{|l|}{Advanced forward-edge techniques for general systems (\textbf{Section~\ref{sec:forwardedge}}):}                                                                                                                                                                                                                                                                                                                                                                                                                                                            \\ \hline
BBB-CFI~\cite{he2020bbb}                                                                                         &                                                                              & \checkmark                                                      & \checkmark                                                                      & \begin{tabular}[c]{@{}l@{}}Block-based enforcement - binary-only approach \\ without need for CFG\end{tabular}                                                                 \\ \hline
PathArmor~\cite{van2015practical}                                                                                       &                               \checkmark                                               & \checkmark                                                      & \checkmark                                                                      & Context-sensitivity - requires architectural support  
                                                          \\ \hline
\multicolumn{5}{|l|}{CFI for embedded systems (\textbf{Section~\ref{sec:cfi_resource_constrained})}}                                                                                                                                                                                                                                                                                                                                                                                                                                                            \\ \hline
Silhouette~\cite{silhouette}   (Section~\ref{sec:advanced_shadow})                                                                                   &                                                                              & \checkmark                                                      & \checkmark                                                                      & Uses shadow-stacks and labeling                                                                                       \\ \hline
\begin{tabular}[c]{@{}l@{}}Control-flow \\ locking~\cite{cflock} \end{tabular}      (Section~\ref{sec:cfl})                           &                  \checkmark                                                            & \checkmark                                                      & \checkmark                                                                      & Lazy + shadow-stack replacement.                                                            \\ \hline
\begin{tabular}[c]{@{}l@{}}$\mu$RAI~\cite{almakhdhubmurai},\\ Zipper Stack~\cite{li2020zipper},\\ PACStack~\cite{liljestrand2020pacstack}\end{tabular}  (Section~\ref{sec:registerbased})                  &                                                                              &                                                                                & \checkmark                                                                      & \begin{tabular}[c]{@{}l@{}} Register-based CFI - shadow-stack replacement \\ Interrupt-handling ($\mu$RAI)        \end{tabular}                                                                 \\ \hline
\begin{tabular}[c]{@{}l@{}}CFI CaRE~\cite{nyman2017cfi}, \\ TZmCFI~\cite{kawada2020tzmcfi}\end{tabular}    (Section~\ref{sec:trustzone})                                 &                                                                              &                                                                                & \checkmark                                                                      & \begin{tabular}[c]{@{}l@{}}ARM TrustZone based shadow-stack, nested \\ interrupts Stronger threat models\end{tabular} \\ \hline
\begin{tabular}[c]{@{}l@{}}HCFI~\cite{christoulakis2016hcfi} (Section~\ref{sec:trustzone})\end{tabular}                                     &                                                                              &                                                                                & \checkmark & \begin{tabular}[c]{@{}l@{}}New ISA that integrates shadow-stack \\ operations in processor pipeline\end{tabular} \\ \hline
\multicolumn{5}{|l|}{CFI for real-time embedded systems (\textbf{Section~\ref{sec:cfi_real-time})}}                                                                                                                                                                                                                                                                                                                                                                                                                                                            \\ \hline
RECFISH~\cite{walls2019control}     (Section~\ref{sec:cfi_rtos})                                                                                     & \checkmark                                                    &                                                                                & \checkmark                                                                      & \begin{tabular}[c]{@{}l@{}}Large-scale schedulability study of common \\ CFI techniques applied to an RTOS\end{tabular}                           \\ \hline
\begin{tabular}[c]{@{}l@{}}Improve schedulability \\ by reducing security~\cite{hao2019integrating}  \end{tabular}           (Section~\ref{sec:exhaustive_search})                                                                               &                                                     &                                                                                & \checkmark                                                                      & \begin{tabular}[c]{@{}l@{}}Searching the number of task jobs that can have \\ CFI turned on to improve schedulability\end{tabular}                                      \\ \hline
  
  Timing-deviation~\cite{bellec2020attack}    (Section~\ref{sec:bellec})                                                                            &                                                                              &                                                                                &                                                                                                & \begin{tabular}[c]{@{}l@{}}Detects control-flow deviation by excess \\ computation time\end{tabular}                                 \\ \hline
  ECFI~\cite{ecfiabasi} (Section~\ref{sec:ecfi})                                                                                           & \checkmark                                                    &           \checkmark                                                                     & \checkmark                                                                      & \begin{tabular}[c]{@{}l@{}}CFI for hard-real time PLC code that detects \\ abnormal increase in execution\end{tabular}                                      \\ \hline

\end{tabular}
\vspace{1em}
\caption{Table of contents of advanced forward-edge CFI techniques discussed in Section~\ref{sec:forwardedge}, CFI techniques for embedded systems discussed in Section~\ref{sec:cfi_resource_constrained}, and CFI techniques for real-time embedded systems discussed in Section~\ref{sec:cfi_real-time}. Important highlights of each technique and degree of coarseness of forward-edge path deviations is discussed.}
\label{tab:table_of_contents}
\end{table*}


  

We shall now look at some general techniques that are used in many CFI mechanisms. We will first look at techniques developed to prevent an attacker from modifying return sequences of function calls (\textit{backward-edge}) or modifying other points-of-interest, such as indirect branches/function calls (\textit{forward-edge}). Techniques for the former are well established and extensively utilized in mechanisms for embedded systems and real-time embedded (Section~\ref{sec:cfi_resource_constrained} and Section~\ref{sec:cfi_real-time}). However, some recently proposed advanced techniques for forward-edge CFI have not yet been considered for real-time embedded systems and are highlighted in Table~\ref{tab:table_of_contents}. \red{Note that for this section and the rest of this paper, ``performance overhead" and ``overhead", unless stated otherwise, are synonymous and refer to the increase in the CPU cycles required due to the addition of the CFI mechanism into the system.``Memory overhead" refers to the increase in the total memory (code and data) required to implement the mechanism, unless otherwise specified. Unfortunately, not all prior work discussed in this survey utilized the same benchmarking software and hardware. Neither did they always report memory overheads. We present the information regarding overheads as it was presented in the original work. We only quantitatively compare different work if the overheads have been measured using the same combination of hardware and software. That said, we try to provide a qualitative discussion when possible to aid the reader in determining the pros/cons of the CFI technique based on the values reported.}

\subsection{Backward-edge CFI techniques}
\label{sec:shadowstackandcanaries}

\begin{figure}[t]
\includegraphics[width=\linewidth, page=1]{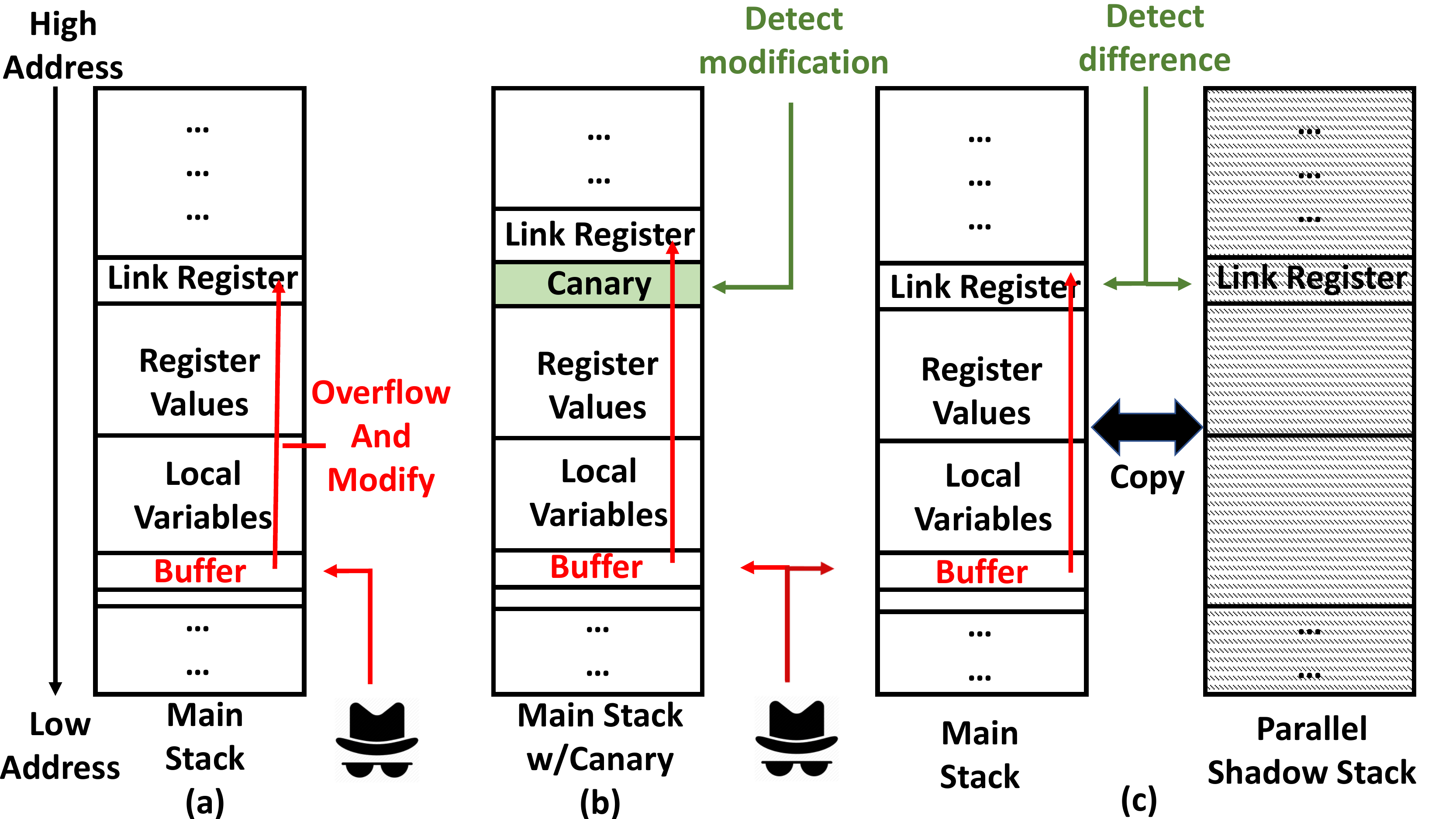}
\caption{(a) A function stack without any defenses, (b) Backward-edge CFI using an embedded canary, or (c) a parallel shadow stack (Section~\ref{sec:shadowstackandcanaries}).}
\label{fig:canariesshadow}
\end{figure}

The first step to any control-flow attack is infiltration. There must be some flaw in the system that can be exploited by an external attacker to begin a control-flow redirection. A very common software flaw is the buffer overflow. Due to the tight memory restrictions of embedded systems, and the flat memory model due to the lack of complex (and potentially expensive, from both economic and performance perspectives) memory management units, buffer overflow or stack overflow flaws are common in resource-constrained embedded systems since they are usually programmed using memory-unsafe languages such as C/C++~\cite{bufferinc}. A simple example of such a flaw is a statically allocated array that is filled past its capacity. Imagine such a flaw exists within a function call of a driver code that handles user input from a keyboard. In the absence of proper memory management, such flaws can be easily exploited to overwrite adjacent locations within the function stack frame as seen in Figure~\ref{fig:canariesshadow}(a). Of particular interest is the return address value in the stack frame. Overwriting the return address with a target address ensures that when the function returns, the code will continue execution at the target address, successfully redirecting the flow of the program. The target address could be either a location within the pre-existing code memory or to some other memory address. A simple use case for the latter technique is to first \textit{inject} the malicious code into the stack memory using the overflow vulnerability, and then set the return address to the start of the injected code. When the function returns, the injected code executes. Code-injection attacks can be thwarted with the help of memory protection mechanisms that implement the W $\oplus$ X memory policy, i.e., prevent execution from writable memory. Such memory protections are now readily available in many commercial-off-the-shelf (COTS) low-end processors and microcontrollers. Therefore, the rest of our discussion will be focused on the consequences of the former technique of forcing the processor to continue execution at a target address in code memory. 

Pointing the processor to an incorrect location by overwriting the return address is an example attack that serves as an entry point to a set of very powerful \textit{code-reuse attacks}. For example, a well-studied sub-family of control-flow attacks is Return-Oriented Programming (ROP)~\cite{roemer2012return}. A ROP attack is where an attacker chain together arbitrary code sequences (also called \textit{gadgets}) that are already present on the device to achieve their objective. Post the seminal work by Shacham~\cite{shacham2007geometry}, ROP attacks have become increasingly popular and very sophisticated. It should be noted that using the return address to perform a control-flow diversion is also referred to as \textit{backward-edge} control-flow attack. On the other hand, \textit{forward-edge} control-flow attacks modify function pointers, or the targets of indirect function calls, to reuse code. An example is that by Checkoway et.al.~\cite{checkoway2010return} that modifies the target of indirect function calls to create gadget chains. Forward-edge defenses are discussed in the next section and are slightly more ambiguous in nature. It is interesting to note that all these attacks require exploiting an initial vulnerability such as a simple buffer overflow bug.

Two simple mechanisms to deal with backward-edge control-flow attacks are stack canaries~\cite{cowan1998stackguard} and shadow stacks~\cite{burow2019sok}. Both these mechanisms, especially the latter, feature heavily in more sophisticated realistic CFI mechanisms for resource-constrained embedded systems. Stack canaries are special values inserted into the stack frame and are located in between the return address and the local statically allocated variables as seen in Figure~\ref{fig:canariesshadow}(b). The concept behind using stack canaries is that an attacker overwriting the stack using a buffer overflow will have to first overwrite the canary value before overwriting the return address. Checking the canary value in the stack frame before a return operation can help determine whether the return address can be trusted. However, stack canaries can be bypassed by a sophisticated attacker, especially if the canary value is known (not random) or if the value can be guessed (not random enough). Further, they do not stop the attacker from overwriting local variables located before the canary value. By doing so, the attacker can still influence the function call operation~\cite{richarte2002four}. 

Shadow stacks are a more sophisticated defense mechanism. Under the assumption that the attacker cannot access or modify a portion of the memory, a copy of the stack frames, or at least return addresses, is kept in that memory portion. Figure~\ref{fig:canariesshadow}(c) presents an example of a shadow stack. This copy is updated during the initial stages of a function call (such as in the function prologue), and the return address is checked just before the return instruction is executed. If a discrepancy exists between the stored and actual addresses, it can be indicative of an attack. Shadow stacks are essentially more sophisticated canaries since both mechanisms indicate an attack by checking for discrepancies in the contents of the stack, with the major difference being that the shadow stack keeps a copy of the correct value~\cite{dang2015performance}. While these mechanisms are relatively simple, applying them comes at a cost. 

Dang et.al.~\cite{dang2015performance} performed a study of the overheads caused by two different shadow stack implementations on the SPEC CPU2006~\cite{specbenchmarks} standard suite of benchmarks on an x86 architecture processor. The first is a "traditional" shadow stack that has its own stack pointer and stores only the return addresses. The second is a "parallel" shadow stack that uses the same stack pointer as the main stack, however, the parallel shadow stack is stored at a different base address and records the return addresses while skipping over the other values in the stack frame (Figure~\ref{fig:canariesshadow}(c)). \red{Architecturally, this makes the parallel shadow stack faster than the traditional shadow stack since the same offset can be used for both the main and shadow stacks. The correct entry can be accessed by simply swapping out the contents of the stack base register which can be achieved with a single instruction. On the other hand, a traditional shadow stack would require additional code to maintain the stack as well as at least one extra instruction per operation to increment or decrement the shadow stack pointer for push and pop operations.} Their measurements of the performance overhead shows that traditional shadow stack implementation, on average, introduces a 9.69\% overhead (over a system without shadow stacks) while the parallel shadow stack introduces a 3.51\% overhead. Worst case overheads of both were 52.5\% and 19.6\%, respectively. The cost of checking the return address was an additional 0.8\%. On the other hand, stack canaries had an average performance overhead of 2.54\%. At first glance the parallel shadow stack mechanism is clearly better suited to applications that are performance sensitive. \red{As discussed above, the performance benefits of parallel shadow stacks is expected since accessing the relevant position in the parallel shadow stack only requires swapping the stack base register since both stacks share the same offset whereas multiple operations are required for an equivalent operation on traditional shadow stacks.} However, the traditional shadow stack has its merits for a resource-constrained system with a low amount of memory.

\subsection{Forward-edge CFI techniques}
\label{sec:forwardedge}

Forward-edge control-flow attacks are the logical extension to backward-edge attacks. The increasing popularity of backward-edge defense mechanisms forced attackers to consider other points-of-interest (POI) to redirect control-flow. These POI include indirect branches and indirect function calls via pointers. By attacking the destination of these branches, the attacker could call any arbitrary location without the need for return instructions~\cite{checkoway2010return}. 

\begin{figure}[t]
\includegraphics[width=\linewidth, page=5]{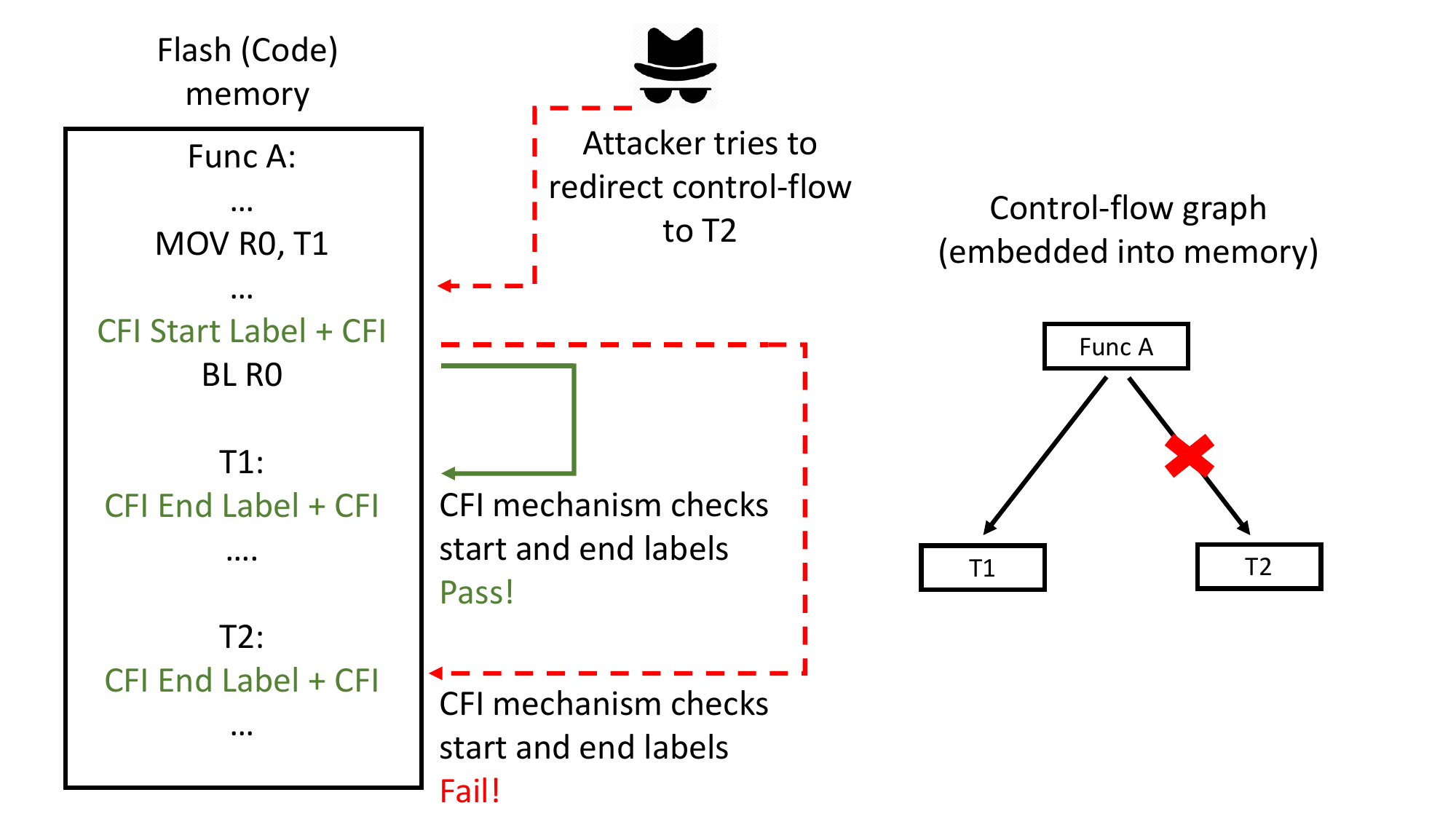}
\caption{Using labels and an embedded control-flow graph to enforce forward-edge CFI (Section~\ref{sec:forwardedge}).}
\label{fig:forwardedge}
\end{figure}

Forward-edge CFI is difficult and, in general, subtler than backward-edge CFI. This is simply because \textit{looking at the past is easier than predicting the future}. Forward-edge CFI techniques that could theoretically predict all possible combinations of branch start and end points are called \textit{fine-grained}~\cite{abadi2009control} CFI. Valid combinations of start and endpoints, essentially valid control-flow, can be represented as a \textit{control-flow graph} (CFG). For example, Abadi et.al.'s~\cite{abadi2009control} approach performs a binary static analysis using Vulcan~\cite{srivastava2001vulcan} to generate a CFG and utilize said CFG to determine whether a branch is valid or not. A common mechanism to help enforce the valid control-flow paths in a CFG is \textit{labeling}. Labeling is a process where all possible forward-edges that can be used by an attacker such as indirect branch locations, functions, and any other potential branch targets, are labeled with unique IDs. Figure~\ref{fig:forwardedge} is an example of a labeling scheme where indirect branches and function prologues are labeled and matched against a CFG. When a branch occurs, the source label (such as an indirect branch) is checked against the destination label (such as a function) via code that has been instrumented into the binary (such as checks in a function prologue). A simple example of such an approach is presented in Figure~\ref{fig:forwardedge}.

An obvious problem of this approach, especially in the resource-constrained embedded systems, is the amount of memory required to store and enforce a CFG. However, more subtle issues arise in real-world cases. Many real-time embedded systems are industrial control systems, robotics systems, etc. In many cases, these environments run proprietary legacy software whose source code is difficult to obtain for analysis, or due to licensing issues, do not allow instrumentation. Due to these reasons, fine-grained CFG may not be possible to obtain, or the performance overhead associated with checking every branch may be prohibitive, especially in a real-time context. Therefore, many \textit{coarse-grained} CFI~\cite{zhang2013control} have been proposed which allow varying degrees of relaxation of which branches or jumps need to be checked and which can be ignored. Due to reduced memory and processing requirements when utilizing coarse-grained CFG, coarse-grained forward-edge CFI are sometimes used for resource-constrained embedded systems. Due to the nature of coarse-grained CFI, such mechanisms may have blind spots that can be exploited by attackers~\cite{coarsegrainbad}. A simple example is where a coarse-grained CFI allows any branch to any legal target, such as the start of a function, due to the unavailability of quality control-flow graphs. In such a case, the attacker could jump to targets which would have otherwise been identified as illegal by a fine-grained CFI. An interesting approach to overcome the need for a CFG, or the codebase to determine a CFG, is proposed by the authors of BBB-CFI~\cite{he2020bbb} where the authors inspect the binary and divide it into \textit{basic-blocks}, with each block having a single entry and exit point. A runtime mechanism prevents branches to the middle of a block, ensuring that the blocks are the smallest unit of code.

Interestingly, even fine-grained CFI can be defeated~\cite{evans2015control, controlflowbend}, such as by exploiting the inability of current code static-analysis techniques to perfectly capture coding practices. Advanced forward-edge CFI techniques such as Van Der Veen et.al's PathArmor~\cite{van2015practical} can defend against such attacks. \red{PathArmor logs control-flow transfers and then performs path verification by having access to the program CFG and performs a depth-first comparison of the logged transfers with the CFG to determine if the path taken during runtime is legitimate. This allows checking if a legitimate pair of source and destination addresses of a control-flow transfer are also contextually correct with respect to neighboring transfer events. However, the requirement for architectural support to record control-flow transfers prevent its direct application to low-end microcontroller-based systems that lack such specialized hardware. }

\subsection{\red{A note on control-flow checking for soft errors}}
\red{
While CFI techniques are built considering an adversarial perspective, there exists a line of research that applies similar methodologies to detect erroneous control-flow redirection due to non-malicious soft-errors~\cite{Zhang2018,Rhisheekesan2019,Gu2014,Schuster2017}. These works utilize very similar techniques, such as by creating signatures for each basic block (code blocks that are delineated by control-flow transfers but do not contain any transfers themselves) and comparing currently executing basic block against a pre-determined graph of valid signature chains~\cite{oh2002control}. While such techniques utilize similar underlying principles to those discussed in prior sections, such as the forward-edge techniques in  Section~\ref{sec:forwardedge}, soft-errors are generally one-shot errors that arise due to environmental factors. Control-flow redirection that is caused due to these errors are not easily predictable. For example, a redirection could take place due to reading the incorrect branch target from memory due to a bit-flip that took place in memory. However, control-flow redirection due to attacker control takes place under more predictable conditions (such as a buffer-overflow bug) and at a control-flow transfer point such as a branch/return statement. Further, advanced control-flow redirection, such as control-flow bending~\cite{controlflowbend}, where a control-flow transfer has valid start and end points but is incorrect only within the context of past control-flow transfers, cannot be detected by control-flow checking techniques since they are built to detect single-shot soft errors. For this survey, we will focus on techniques explicitly built for defending against various forms of control-flow redirection attacks.}

\section{CFI for Embedded Systems}
\label{sec:cfi_resource_constrained}
We now move towards more realistic CFI implementations in the context of resource-constrained embedded systems. The mechanisms presented here either combine techniques from Section~\ref{sec:basic_techniques} or propose entirely new techniques. Highlights of some of the mechanisms discussed in this section are presented in Table~\ref{tab:table_of_contents}.

\subsection{Implementation of basic techniques}
\label{sec:advanced_shadow}

We stated a pre-requisite in the prior section with respect to shadow stacks - \textit{\ldots Under the assumption that the attacker cannot access or modify a portion of the memory}. This assumption does not have a straightforward justification in the context of embedded systems. As previously noted low-end embedded systems simply do not have complex memory management units to support well-known features such as virtual memory, which is now common in higher-end processors, let alone have special built-in mechanisms to support hiding shadow stacks from an attacker. Therefore, a successful CFI mechanism has to first wrangle the available hardware capabilities to support shadow stacks.

Zhou et.al's Silhouette~\cite{silhouette} is an attempt to support shadow stacks on ARMv7-M~\cite{armholdings7m}, the architecture underlying ARM Cortex-M series of processors commonly found in embedded systems. It also supports forward-edge CFI checks. Silhouette, thus, is an example of how a sophisticated CFI mechanism would look like in the context of a resource-constrained embedded system. The ARMv7-M architecture supports two privilege levels in hardware, privileged and unprivileged. The optional memory protection unit (MPU) allows a system designer to decide access rights to an address.  A limitation of the ARMv7-M architecture is that the MPU can be controlled by any privileged code. \red{Most RTOS, such as FreeRTOS~\cite{barry2008freertos}, by default, execute both the tasks and the operating system as privileged code to mitigate the overhead of switching privilege levels. This makes using the MPU to protect a shadow stack a moot point, simply because an attacker that has infiltrated the system, could re-program the MPU since they would most likely already execute under the privileged execution context. }

Silhouette ensures that the MPU access rights are adhered to by working around this limitation. It replaces all store instructions, other than those that are supposed to directly store to the shadow stack, or the hardware abstraction layer (HAL) code, with unprivileged store variants, at compile time, to ensure adherence to the memory access policies defined in the MPU for the target address, regardless of the processor's current execution privilege level. The shadow stack is implemented in a similar manner as the parallel shadow stack explained in Section~\ref{sec:shadowstackandcanaries}. To ensure that the store instructions with higher privilege levels are not abused by an attacker, Silhouette implements forward-edge CFI checks. Silhouette utilizes a labeling mechanism (Section~\ref{sec:forwardedge}) to guarantee forward-edge CFI~\cite{cfiprecision}.

On the performance front, Silhouette is benchmarked using well known embedded systems benchmark suites, namely CoreMark-Pro~\cite{eembc2015coremark} and BEEBS~\cite{pallister2013beebs}. We will see these same benchmarks being used in other approaches too in later sections, providing a common playing field. The maximum performance overhead reported for the two benchmark suites is 4.9\% and 24.8\%, respectively, and a code memory overhead of 8.9\% and 2.3\% respectively. The geometric mean of the performance overheads for all the benchmarks in each test suite is 1.3\% and 3.4\%, respectively. The approach used by Silhouette, which they term as \textit{store hardening}, basically utilizes a memory management technique to hide the shadow stack from the attacker. 

Another mechanism that can be used to prevent access to the shadow stack is called software fault isolation~\cite{wahbe1993efficient,mccamant2005efficient} (SFI). SFI is a technique where the address space is partitioned into \textit{fault domains}. Any code within a fault domain has unrestricted access to code or data within the same fault domain, but the partitioning scheme prevents the code from accessing any memory outside the fault domain. \red{This is achieved by instrumenting load/store instructions during compile time to trigger the fault handler if the memory access takes place outside the fault domain.} A variant of Silhouette is proposed that utilizes this technique \red{by instrumenting store instructions to restrict them from writing to the shadow stack unless the store instruction is part of the shadow stack manipulation code}. The authors note a higher performance overhead, with the geometric mean results being 2.2\% and 10.2\% respectively for the two benchmarks, which leads the authors to conclude that the store hardening approach is superior in performance. However, it would be interesting to note how the performance would vary if the shadow stack was protected using an approach similar to Aweke and Austin's~\cite{usfi} lightweight SFI for IoT systems that shows an overhead of just 1\% on the MiBench~\cite{guthaus2001mibench} benchmarks. Their approach utilizes a small amount (150 lines) of trusted code that sets up the MPU to create the fault domains, trapping accesses outside the domain as memory access faults. Unfortunately, they do not present results using the CoreMark-Pro or BEEBS suites making direct comparisons difficult. 

While the Silhouette and its variant provide a good overview of the well-known techniques of shadow-stacks and labels can be applied to a real low-end processor architecture, there are avenues to improve the operation of such systems. 


\subsection{Beyond the basics}
\label{sec:cfl}

\begin{figure}[t]
\includegraphics[width=\linewidth, page=2]{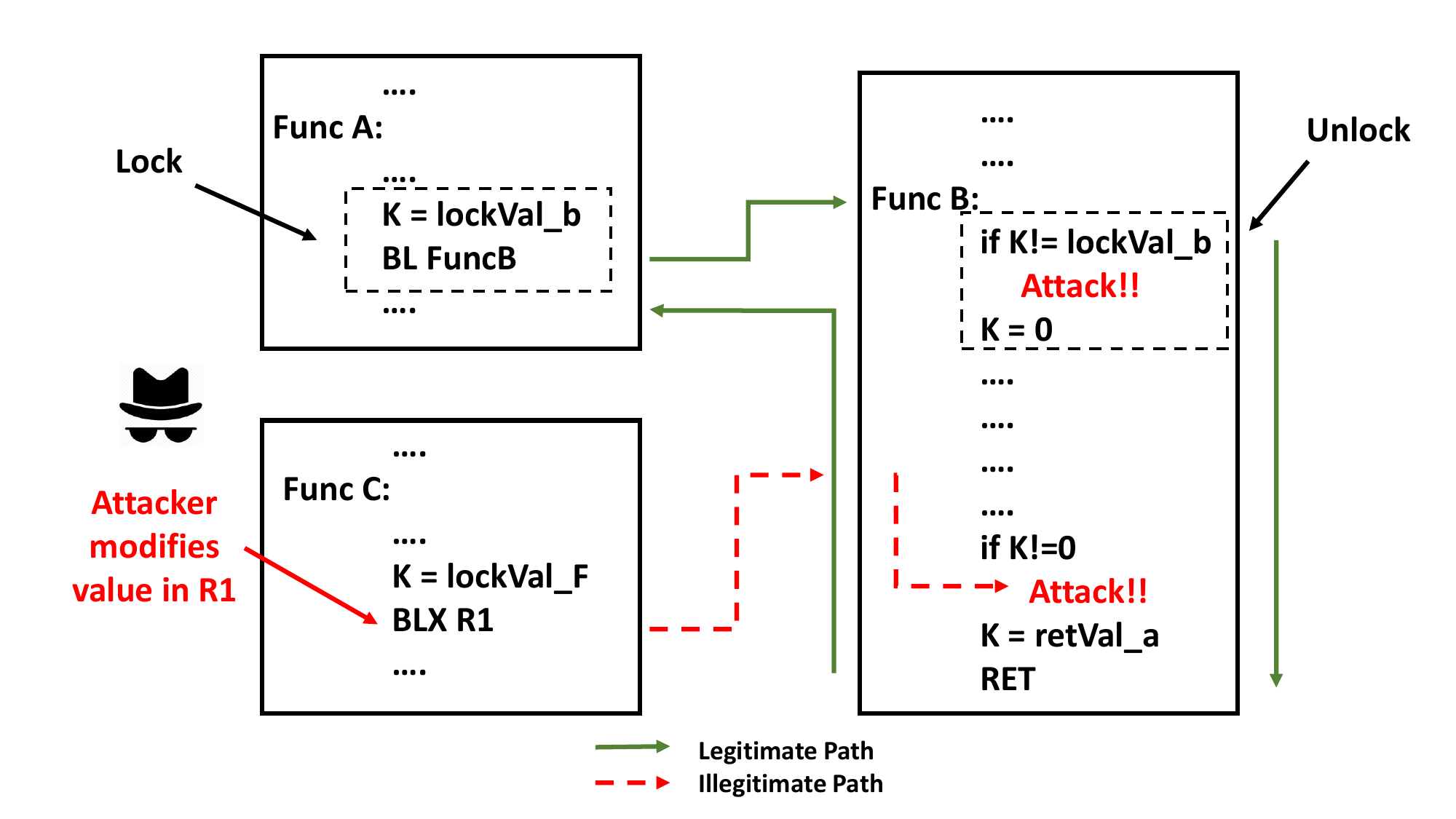}
\caption{Control-flow locking operation. Note the exclusive use of lock/unlocks for the entire operation (Section~\ref{sec:cfl}).}
\label{fig:controlflowlock}
\end{figure}

While the techniques discussed in Section~\ref{sec:basic_techniques} consider forward-edge and backward-edge separately, some effort has been applied in recent years to develop more holistic mechanisms that apply to backward and forward-edges at the same time.

An example of such a mechanism is the Control-Flow Locking (CFL) technique~\cite{cflock}. This is also an example of a \textit{lazy} CFI that trades-off attack detection speed with performance overhead. While CFL is not explicitly targeted at resource-constrained embedded systems, the mechanism can be implemented with similar memory and performance overhead as any general label-based CFI for detecting forward-edge control-flow attacks. CFL uses locks, instead of shadow stacks, to determine if an attacker has diverted control-flow to an arbitrary location. An overview of the CFL operation is given in Figure~\ref{fig:controlflowlock}.  The idea behind CFL approach is simple. Similar to how labels are generated based on the valid control-flow graph, \textit{key values} are assigned to legitimate call/jump target locations. CFL targets indirect calls/jumps as well as return instructions (an x86 architecture-based processor was assumed). Once the unique key values, which essentially represent valid edges in the control-flow graph, are generated, the authors propose to then instrument the target binary with instructions to lock and unlock control-flow paths using these key values. Every legitimate control-flow redirection start point, which may be an indirect \texttt{call, jmp} or \texttt{ret} instruction, is \textit{preceded by a lock operation}, i.e., the key value is stored into a buffer. The assumption here is that the buffer is stored in a memory location such that it can be modified only by the lock and unlock subroutines, and not by attacker-controlled code. Once program control is redirected to a valid destination (such as a function entry-point), it is \textit{immediately succeeded by an unlock operation} where the key value is validated, i.e., it is checked against a list of key values that could end up at this target location. If the values match, the key is zeroed out (\textit{unlocked}) and execution continues as before. When the next control-flow redirection operation must take place, the key buffer is first checked to see if it contains a non-zero value. If it does, an attack is detected since no legitimate transfer would allow the key buffer to have a non-zero value due to the paired lock-unlock operations. Depending on the quality of the available CFG, this pairing of lock-unlock operations could be coarse or fine.


The overall mechanism is interesting due to its simplicity and the introduction of laziness. Not only does it prevent an illegitimate jump to a \textit{valid} control-flow transfer site, but also automatically detects an illegitimate jump to an \textit{invalid} control-flow site in recent history \textit{without} requiring additional runtime memory such as using a shadow stack. Evaluations show that CFL can outperform fine-grained CFI mechanisms , with a maximum overhead of 21\% v/s 31\% overhead under Abadi et.al's~\cite{abadi2009control} mechanism on the SPEC CPU2000~\cite{specbenchmarks} benchmarks. The tradeoff? - As mentioned earlier the mechanism is lazy. The attacker can redirect control and can remain undetected until it is caught by the next locking site. While laziness allows the mechanism to work with the time and memory overhead similar to a labeling scheme, it could have interesting security repercussions especially in the context of the real-time embedded systems, many of which are used in industrial environments, controlling actuators in critical processes. If an attacker is able to send out control commands to these actuators before they are detected, the attacker can still inflict catastrophic damage. However, laziness is not inherently flawed. There is therefore an avenue to leverage real-time requirements to enforce timing bounds on laziness. 

While CFL is an example of a CFI technique that re-purposes control-flow labels to solve both forward and backward control-flow attack detection at the same time, it still uses a form of memory protection. All the techniques discussed up to this point attempt to work around hardware limitations to enforce memory protection and are conservative. However, they do not take full advantage of the processor architecture or require radical software/hardware changes to improve performance. 

\subsection{Register-based shadow stacks } 
\label{sec:registerbased}
We will now discuss two approaches that would require significant software modifications to allow them to work. We will first briefly look at Zipper Stack~\cite{li2020zipper} which is the more radical of the two since it proposes CPU architecture modifications to forego shadow stacks. The other is $\mu$RAI~\cite{almakhdhubmurai} that is built for COTS embedded systems. It takes a more moderate approach by requiring reservation of parts of the CPU but can be implemented by recompiling the codebase with a modified compiler. Both implement backward-edge CFI.

\red{Zipper stack aims to solve the problem of securing shadow stacks by replacing them with a set of processor architecture modifications. Shadow stacks, as discussed in Section~\ref{sec:basic_techniques} are inherently simple but require additional support to secure them from attacker manipulation. For example, Silhouette in Section~\ref{sec:advanced_shadow} requires additional code instrumentation to secure the shadow stack}
Zipper Stack aims to solve this problem by replacing the shadow stack with a single value stored in a special-purpose register called the \textit{top register}. A separate register, the \textit{key register}, holds a secret key. At the start of a new process, the key register and top register are initialized with random values. Each time a function call takes place, the top register is pushed onto the main stack alongside the actual return address. A message authentication code (MAC) algorithm, a cryptographic operation that is commonly used to authenticate messages from a known source, generates a new MAC from the top register value and the return address using the key in the key register. This newly created MAC is then stored in the top register. During a return sequence, the steps are reversed to authenticate the return address. First, the previous MAC value is popped from the stack and the MAC is recalculated using the return address and the popped MAC value. If the calculated MAC matches that currently in the top register, the return address is verified to be authentic. The processor replaces the top register with the popped value and continues execution at the return address. \red{The purpose of the MAC based design is to reduce the attack surface. By utilizing the top register and chaining the MAC values with each successive function call, an attacker can only modify the return address and evade detection if it first modifies the value present in the top register (which is inaccessible to application code and is automatically updated by the hardware) before modifying the other MACs. Therefore, the rest of the MACs can be kept in non-secure memory that may be accessible to the attacker, reducing the amount of overhead introduced by accessing the ``zipper stack" of MAC addresses.}

The operation shows that Zipper Stack is heavily dependent on a) the efficacy of the MAC algorithm to ensure \textit{collisions} (same MAC from different inputs) do not occur, b) the speed of the algorithm since every function call would constitute running the algorithm at least twice, and c) the attacker not being able to access the key register to forge MACs. For a), The authors use a well-known MAC algorithm, for b) the authors argue that a hardware implementation would allow MAC calculation in a single cycle, and for c) the authors argue that even if the key is leaked, the top register can only be modified at a call or a return operation. Their custom implementation on an FPGA with a RISC-V CPU achieves a 1.86\% overhead on the SPEC CINT 2000~\cite{specbenchmarks} benchmark. 

While Zipper Stack presents a very radical approach that may never see wide-scale commercial adoption due to its hardware modifications, it is still interesting since custom architectures for specific applications, such as defense, are not uncommon in the embedded system world. In such cases, a custom architecture designed with optimized built-in defense mechanisms is not hard to envision. Interestingly, the use of MACs for authenticating return address may become possible very soon on commodity hardware. For example, PACStack~\cite{liljestrand2020pacstack} re-purposes the ARM \texttt{pac} instruction to create a MAC chain of return addresses, very similar to Zipper Stack. As part of the ARMv8.3-A PA extension, and soon to be available on SoCs based on ARMv8.3-A and later architecture revisions, \texttt{pac} allows generating pointer authentication codes (PAC) which are MACs generated on pointer values and stored along side the pointer. Similar to Zipper Stacks, the authors use a \textit{chain register} to store PAC values which are generated from previous chain register values and the return address of a function call. When a return sequence takes place, similar to Zipper Stack, the reverse operation takes place. PACStack showed a geometric mean of 2.75\% and 3.28\% performance overhead on the SPECrate and SPECspeed (part of the SPEC CPU 2017 benchmark suite), respectively. PACStack provides a strong argument for MAC based shadow stack replacement, especially since it depends on architecture extensions which will soon be available in commodity hardware. 

On the other hand, the authors of $\mu$RAI take a similar but more realistic approach, especially on current-generation hardware. $\mu$RAI is also concerned solely with the backward-edge, but instead of verifying the return address as is common with shadow stack approaches, $\mu$RAI enforces Return Address Integrity (RAI) where the return address simply cannot be modified by an attacker. Their approach, in essence, is to prevent write access to the return address. $\mu$RAI has the same set of requirements as many of the schemes we have discussed in previous sections, such as data execution prevention (DEP or $W \oplus X$) and an MPU. Similar to Zipper Stack, it requires that one of the processor registers is wholly dedicated to its operation and should never spill. This is called the \textit{State Register} (SR). $\mu$RAI's operation requires that the attacker cannot modify the register. 

\begin{figure}[t]
\includegraphics[width=\linewidth, page=3]{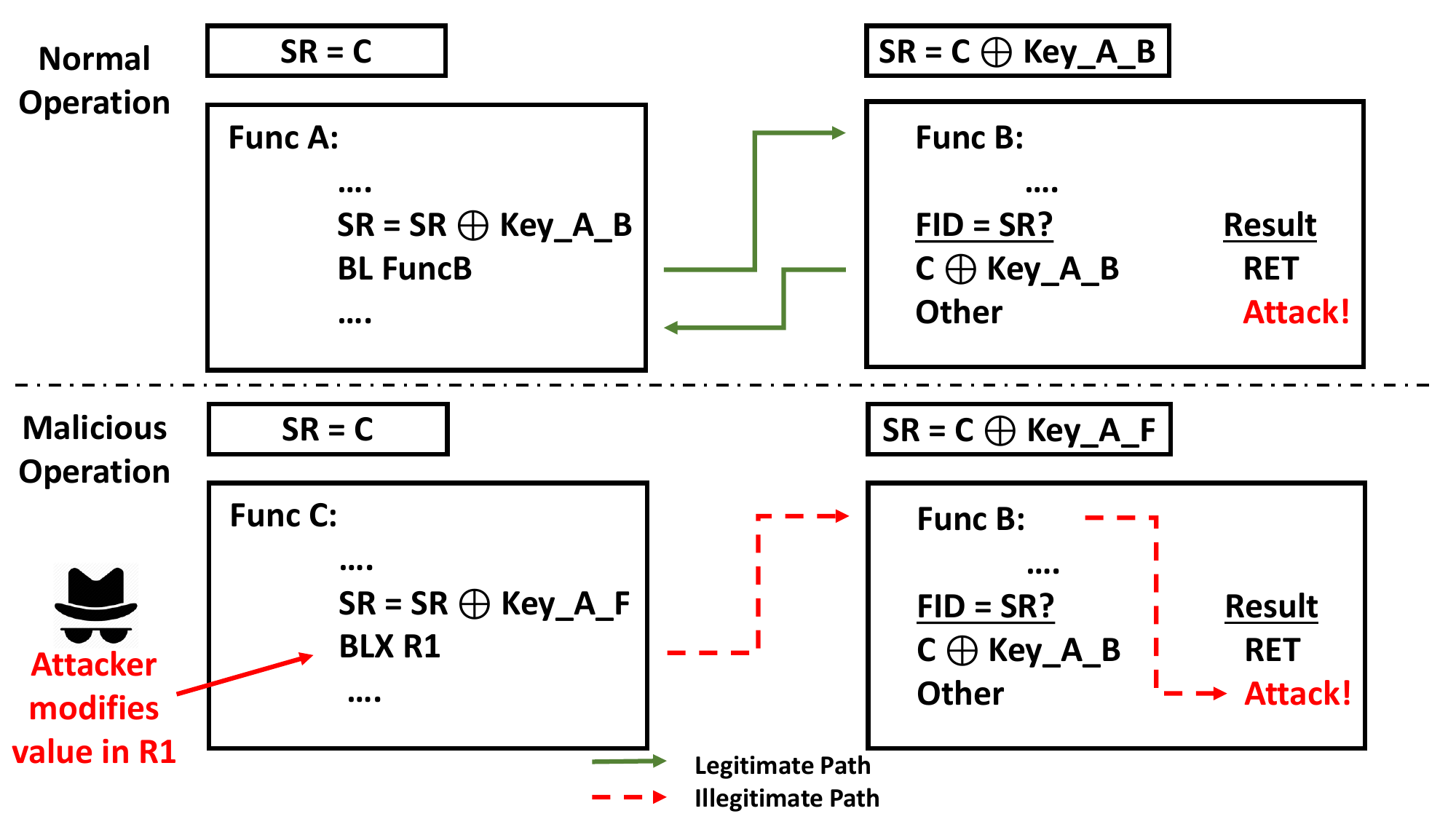}
\caption{$\mu$RAI operation. Shadow stack operation is implemented via SR register and FID table during return (Section~\ref{sec:registerbased}).}
\label{fig:muRAI}
\end{figure}

$\mu$RAI works by instrumenting code before branches and at return points, similar to CFL. It works solely with direct branches, i.e. branches with encoded destinations, and converts all indirect branches into direct branches by matching all possible start and endpoints. Figure~\ref{fig:muRAI} provides a basic overview of how $\mu$RAI instrumented code looks like and operates. Every function \textit{call site} is assigned a unique function key (FK). As is seen in the figure, a Function A can have multiple call sites to another Function B. $\mu$RAI instruments code such that before every such call site, the value in the SR register is XOR'ed with the FK for the call site. This value is also called the Function ID (FID). The call goes through and Function B operates. At the point where Function B returns, it checks what the authors call the Function Lookup Table (FLT). This table has all the FIDs that could call this function. Based on which FID matches the value in the SR, the function returns to the corresponding location. Finally, the SR is XOR'ed with the same FK used before the branch, returning it to the original value before the function call. The authors tested their approach on an ARM Cortex-M4 based board and report a maximum performance overhead of 8.1\% on the CoreMark~\cite{gal2012exploring} (a lighter variant of CoreMark-Pro) benchmark with an average of just 0.1\%, making it comparable with shadow stack mechanisms discussed previously. However, it requires on average 34.6\% extra flash memory for instrumentation and FLT. 

The reader may have noticed that the possible return addresses are encoded into the code memory under DEP restrictions that prevent an attacker from modifying the code memory. DEP is enforced using the MPU. $\mu$RAI, therefore, foregoes the return address that the processor may record in its stack, which is inherently writable memory, during a function call. Instead, it implements a function return mechanism that is implemented completely in code memory. This enforces $\mu$RAI's goal of return address integrity. $\mu$RAI is also the first mechanism that we have discussed in this survey, that explicitly considers interrupts. Since interrupts can occur at any time and can potentially interfere with shadow stack operations, they require explicit consideration. $\mu$RAI instruments interrupt handler code to first save the return address which has been automatically stored on the stack by the hardware before the handler code is executed. $\mu$RAI saves the return address to a safe memory hidden behind the MPU. Here $\mu$RAI has to essentially create a shadow stack due to the limitation of the hardware. Supporting interrupts is a significant step to eventually supporting multi-threaded scheduling under a real-time operating system (RTOS). However, dedicating a register to $\mu$RAI operations would require modifications to the compiler as well as incompatibility with embedded systems having a severely limited processing capacity, especially when the software requires large number of registers for computational purposes.

Unfortunately, none of these techniques improve forward-edge CFI. For example, in the case of $\mu$RAI, the attacker could keep redirecting code execution using branch operations without allowing code to execute till an FID table. Therefore, such CFI mechanisms are helpful from only a performance or memory perspective over a regular shadow stack. That is, they do not provide any additional security guarantees, while requiring significant codebase changes or at least a modified compiler to support their operation.   

\subsection{CFI using processor architecture extensions}
\label{sec:trustzone}
 Before we finally move towards real-time aware CFI mechanisms, we will look at two mechanisms that depend on very modern processor architecture extensions such as ARM TrustZone~\cite{pinto2019demystifying}. TrustZone allows a processor to support two execution domains, \textit{secure} and \textit{non-secure}, each with its own address space with the secure domain having supervisory access to the non-secure domain. CFI designers have found creative ways to use it as part of their designs.  

\begin{figure}[t]
\includegraphics[width=\linewidth, page=4]{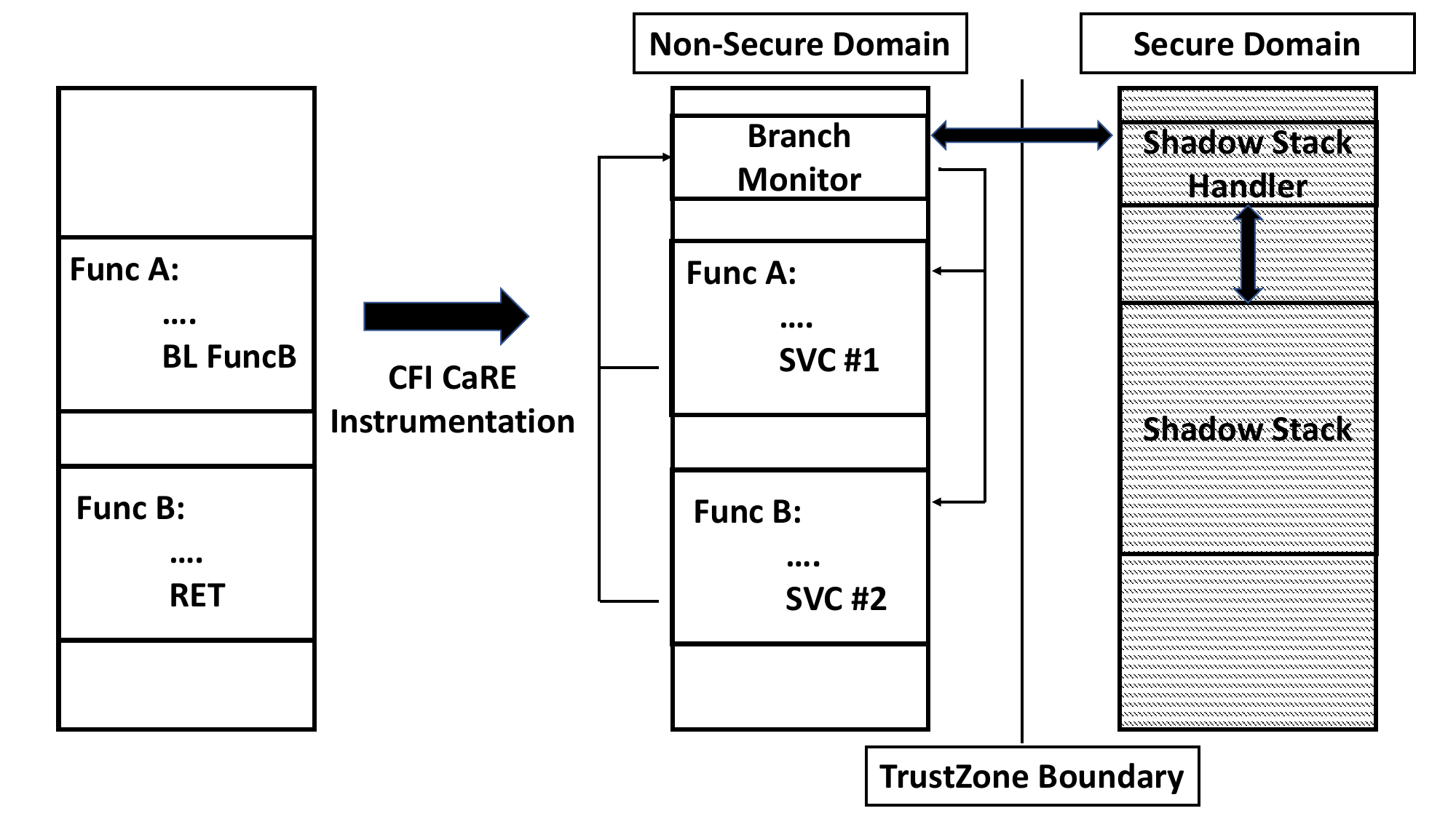}
\caption{CFI CaRE instrumented code. Branch and return instructions translate to SVCs (Section~\ref{sec:trustzone}).}
\label{fig:CFICaRe}
\end{figure}

The first is Nyman et.al.'s CFI CaRE~\cite{nyman2017cfi} that presents an alternative approach to secure the shadow stack, to that of Silhouette~\ref{sec:advanced_shadow}. An overview of its operation is given in Figure~\ref{fig:CFICaRe}. While Silhouette uses binary instrumentation to prevent a privileged attacker from modifying the MPU that hides the shadow stack, CFI CaRE hides the shadow stack behind the TrustZone in the secure domain. CFI CaRE assumes that the original binary is only allowed to execute under the non-secure domain. It replaces all function calls with a \textit{supervisory call} (SVC) that launches a special function called the \textit{branch monitor}. The branch monitor runs in a privileged context and based on the parameter passed to the SVC that launches it, the branch monitor is able to identify if the source of the SVC is a branch or a return. It then calls secure domain code, passing the source identifier as a parameter, that updates the shadow stack. While the SVC ensures that all branches and returns are effectively trapped into the branch monitor, the TrustZone boundary ensures that non-secure domain code cannot view or modify the shadow stack. The authors used the Dhrystone (precursor to CoreMark) benchmarks to evaluate their work on an ARM Cortex-M23 processor. Performance overhead ranged between 13\% to 513\% with an overall 14.5\% increase in flash memory consumption.

While CFI CaRE may seem like just a different implementation from previous approaches, it proposes a mechanism to address a crucial flaw in previous approaches with respect to embedded systems. The previously discussed approaches instrument binaries with no regard to the original layout. While this may be a non-issue for systems whose source code is available, many real-time embedded systems use proprietary legacy software and access to the source code may be limited. Further, due to memory and processor restrictions, these binaries are painstakingly built with strict adherence to page limits, available flash memory, etc. Unchecked binary instrumentation may destroy compatibility with the hardware. CFI CaRE's usage of SVC simply overwrites the branch or return instructions, keeping the original binary layout intact. However, it does require extra space for the branch monitor. 

CFI CaRE also support interrupts and uses \textit{trampolines} which are short sequences of code at the start of interrupt that call the secure domain to store the return address in a shadow stack. However, it does not support nested interrupts. If an attacker-controlled higher priority interrupt fires before the trampoline can store the return address in the shadow stack, the attacker-controlled interrupt code could rewrite the return address. When the lower priority interrupt finally gets to run, its trampoline would store a modified return address. Furthermore, nested interrupts can occur on an RTOS controlled system. For example, the timer tick could fire alongside interrupts from other peripherals. Kawada et.al.'s~\cite{kawada2020tzmcfi} TZmCFI fills in this gap. They too propose using the TrustZone to hide the shadow stack. However, they also extend the shadow stack concept to what they term as \textit{exception shadow stacks} that support nested interrupts. They modify the trampolines such that every trampoline will complete all pending shadow stack transactions of lower priority interrupts before the interrupt body is allowed to execute. This ensures that if an attacker controls the interrupt body, it cannot affect the shadow stack copy of the interrupt return address. TZmCFI showed a performance overhead of up to 84\% when supporting FreeRTOS as compared to FreeRTOS without CFI. For nested interrupts, the instrumented interrupts (with the trampolines) increased interrupt execution time from ~30 cycles (un-instrumented) to 132-236 cycles, i.e., up to a ~550\% increase in execution time. 

\red{Other work that involves extending the architecture of the processing environment include work such as HCFI~\cite{christoulakis2016hcfi} suggest creating a new CFI enabled instruction set architecture (ISA) by modifying an existing ISA such as SparcV8's Leon3~\cite{gaisler2001leon}. They do so by adding new stages in the CPU pipeline to perform CFI operations such as shadow stack operations and show that performance overhead with respect to an umodified Leon3 core is less than 1\% on their FPGA implementation for the SpecInt2000 benchmarks. While optimum performance can be achieved by designing a custom processor core as suggested here, unlike the TrustZone-based approaches discussed earlier, this would require significant investment to implement in real systems in the near future.}

\subsection{\red{CFI using Separate Processing Environments}}
\label{sec:off_chip}

\red{We wrap up our discussion of different CFI mechanisms for embedded systems with a brief note about CFI by utilizing off-chip processing environments since they behave very similarly to CFI achieved via TrustZone and utilize the same set of techniques presented in detail in Section~\ref{sec:basic_techniques}. For example, techniques such as Abad et al.'s~\cite{abad2013chip} uses a separate monitoring module to track the program counter and detect deviation from the control-flow. Similarly, SecMonQ~\cite{nasser2020secmonq} is designed for automotive systems and utilizes the Hardware Security Module (HSM) found in many commercial automotive ECU's to detect anomalous path behaviour. In a more general sense, techniques such as RTTV~\cite{yang2018rttv} utilize the Trusted Platform Module (TPM), a common co-processing environment used as a store for keys for cryptographic keys and perform a limited and static set of cryptographic operations in many embedded systems, can be used to store the CFG and perform regular measurements against the stored CFG. All these techniques inherit and apply the basic techniques presented in Section~\ref{sec:basic_techniques}.}

\subsection{\red{Section Summary}}
\label{sec:resource_constrained_summary}

\red{The techniques discussed in this section generally follow the basic techniques listed in Section~\ref{sec:basic_techniques}. The proposed mechanisms either directly apply those basic techniques, or have progressively complex hardware modifications, from special registers to reduce the cost of shadow stacks (Section~\ref{sec:registerbased}) to novel ISA (Section~\ref{sec:trustzone}). However, the techniques do not inherently change the underlying principles of CFI and can be \textit{conventional} by their nature. That is, they all verify the source and target destination addresses without much variation. Another important observation is that each of the techniques presented is uniquely tied to the underlying hardware for both performance and enforcement of CFI, making it difficult to compare their individual overheads. However, on a qualitative note, it is clear that the most performant CFI require radical hardware changes, such as integrating shadow stack operations into the pipeline of the processor~\cite{christoulakis2016hcfi}. }

\red{A common theme in the techniques discussed, however, is the lack of any discussion regarding the implications of the overhead they introduce on systems where timing is critical, e.g. real-time systems. Real-time systems have certain characteristics that could be utilized to aid CFI and/or reduce the impact of the overhead introduced. We will now discuss these characteristics:
\begin{enumerate}
  \item In periodic real-time systems, work is performed in a temporally predictable manner. That is, \textit{tasks} execute during defined periodic intervals. CFI could utilize this predictable periodic nature to determine if an application is misbehaving due to attacker control. 
  \item The system is usually underutilized due to safety requirements. Since real-time systems are, in many cases, deployed in critical environments such as medical, industrial or automotive systems, such systems are designed to not perform work all the time to reduce or eliminate the possibility of missing deadlines. For example, the system is usually provisioned with enough computing resources such that tasks do not need to consume 100\% of the computing resource at all times to complete by their deadlines. Therefore the system may have large periods of \textit{slack} where the system idles, interspersed with heavy computation phases. CFI could utilize the slack thereby reducing localized spikes in computational load and reducing the possibility of missing deadlines. Note that although these systems may be underutilized, they are still considered to be resource-constrained. The underutilization is intentional due to safety concerns and any addition in the computational requirements must be done judiciously. 
  \item The total system utilization at any given point of time is usually well characterized and there exist schedulability tests to determine if the system may be successfully scheduled without missing deadlines under a given scheduling algorithm and utilization. These tests may differ for different type of real-time task models (periodic tasks, aperiodic tasks, etc.). None of the techniques discuss their applicability and/or changes that must be introduced to satisfy these schedulability tests.
\end{enumerate}
}
\red{However, none of the techniques discussed in Section~\ref{sec:cfi_resource_constrained} consider timeliness, We now discuss CFI work that are specific to real-time embedded systems. 
}

\section{CFI for Real-Time Embedded Systems}
\label{sec:cfi_real-time}

We have discussed multiple CFI techniques in the previous section for embedded systems. In this section, we survey the state-of-the-art mechanisms that consider real-time requirements. Unfortunately, there is little prior work that explicitly consider real-time properties of the system's operation. Therefore, this section discusses a few available CFI mechanisms. We divide our discussion into two parts, the first part covers techniques that are built specifically with an RTOS scheduler in mind, and the second discusses non-conventional CFI approaches. Highlights of the mechanisms discussed in this section are presented in Table~\ref{tab:table_of_contents}.


\subsection{CFI with an RTOS}

\subsubsection{An analytical approach for common CFI techniques}
\label{sec:cfi_rtos}
TZmCFI, presented in the previous section, is an example of CFI mechanisms for embedded systems that can work alongside an RTOS, or more specifically, a scheduler. A scheduler consists of supervisory code that decides when code that does actual work, i.e. complete the goal of the system, is able to run. A scheduler is critical to ensure system timeliness. While TZmCFI supports an RTOS, it lacks a study of system schedulability under different workloads. The recent work by Walls et.al.~\cite{walls2019control} addresses this deficiency in research. Their approach, called \textit{RECFISH}, is an RTOS-aware CFI scheme. Since RECFISH shares several similarities with techniques discussed in prior sections, we will briefly discuss the mechanism and take a closer look at the evaluation results. 

RECFISH is designed for ARM Cortex-R~\cite{cortexr} processors that are built specifically for critical real-time applications. Like the Cortex-M series, they forego memory management units, and have special caching mechanisms to maintain predictability, support a small address space, but do not support TrustZone. RECFISH, instead, utilizes the MPU, like $\mu$RAI, to enforce DEP. It assumes that the task code executes in the unprivileged mode while the RTOS runs in privileged mode. This ensures that if an attacker infiltrates a task, it cannot override the MPU settings. RECFISH is designed to be used with FreeRTOS and modifies it to allow setting up a per-task shadow stack (which only privileged code, such as the RTOS, can modify since it is hidden by the MPU), and modifies the scheduler to update the shadow stack when switching between tasks. Finally, RECFISH also instruments the binary to add labels to function prologues, as well as enforce shadow stack operations before (and after, in the function epilogue) the function body can execute. The labeling mechanism is used for enforcing forward-edge schemes, while the shadow stack operations are enforced by calling privileged shadow stack handling code using SVC just like that seen in CFI CaRE. 

While the operation of RECFISH may look similar to multiple ideas presented in previous sections, the authors are the first to present a study of their approach's effect on real-time workloads. They evaluate and note a 21\% performance overhead for their approach on the CoreMark benchmarks. Microbenchmarks show that RECFISH increases scheduler context switching time from 120 CPU cycles to 159. Further, the label checking and shadow stack operations increase function prologue and epilogue overheads from 19 cycles (without any CFI operation) to 275 cycles. The authors then perform a large-scale schedulability study on simulated workloads. They randomly generated synthetic task sets with varying utilization values, task periods, and number of indirect branches. Utilization values ranging from 0.1\% to 90\% were considered. The overhead of task context switch (39 cycles) was incorporated into the task's worst-case execution time (WCET). For incorporating the function prologue and epilogue overheads (label checking for forward-edge CFI), the authors considered a varying number of indirect branches per task that were either 0, or ranging from 1 every $10^3-10^5$ cycles to 1 every $10^6 - 10^7$ cycles. Multiplying the number of branches with the 256 cycle overhead for the task yielded the overhead for the label checking mechanism which was then incorporated into the task WCET. RECFISH performs well for task sets where the number of tasks is few and each task has a high utilization, and when indirect branches are infrequent. However, the results show that up to 30\% of the system utilization can become unusable for task sets with more frequent indirect branches and function calls and more tasks. Overall, RECFISH could schedule 85\% of the 6 million task sets generated from 5760 different parameter combinations. The results show that well known CFI mechanisms such as shadow-stack and labeling could be used with a wide range of multi-threaded real-time workloads. 

\subsubsection{\red{Trade-off security for schedulability}}
\label{sec:exhaustive_search}

\red{While RECFISH provides a schedulability study of common CFI techniques, Hao et al.~\cite{hao2019integrating} provides a novel technique to improve the schedulability of a real-time system by trading-off security with system schedulability. They focus on defending against ROP attacks (Section~\ref{sec:introduction}). They do so by selectively switching on CFI checks for a subset of instances (also called jobs) for each task in the system by exhaustively searching for the maximum set of jobs that can have CFI checks without hindering the schedulability of the system. The authors provide a comparison of an approximated scheduling algorithm that is designed to be faster to execute during runtime with respect to the exhaustive search algorithm which is determined to be optimal. Experimental results show that their approximation approaches optimality at lower ($\leq 0.6$) utilizations. A schedulability study shows that there is a sharp drop-off in schedulability of task sets if the CFI checks are added to task sets with utilization greater than 0.8. This observation echoes the results of the study of RECFISH that as task sets become "heavier", that is, have a higher utilization, schedulability sharply drops down to zero.  
}
\subsection{CFI utilizing timing deviations}
\label{sec:timing_deviations}

\begin{figure}[t]
\includegraphics[width=\linewidth, page=6]{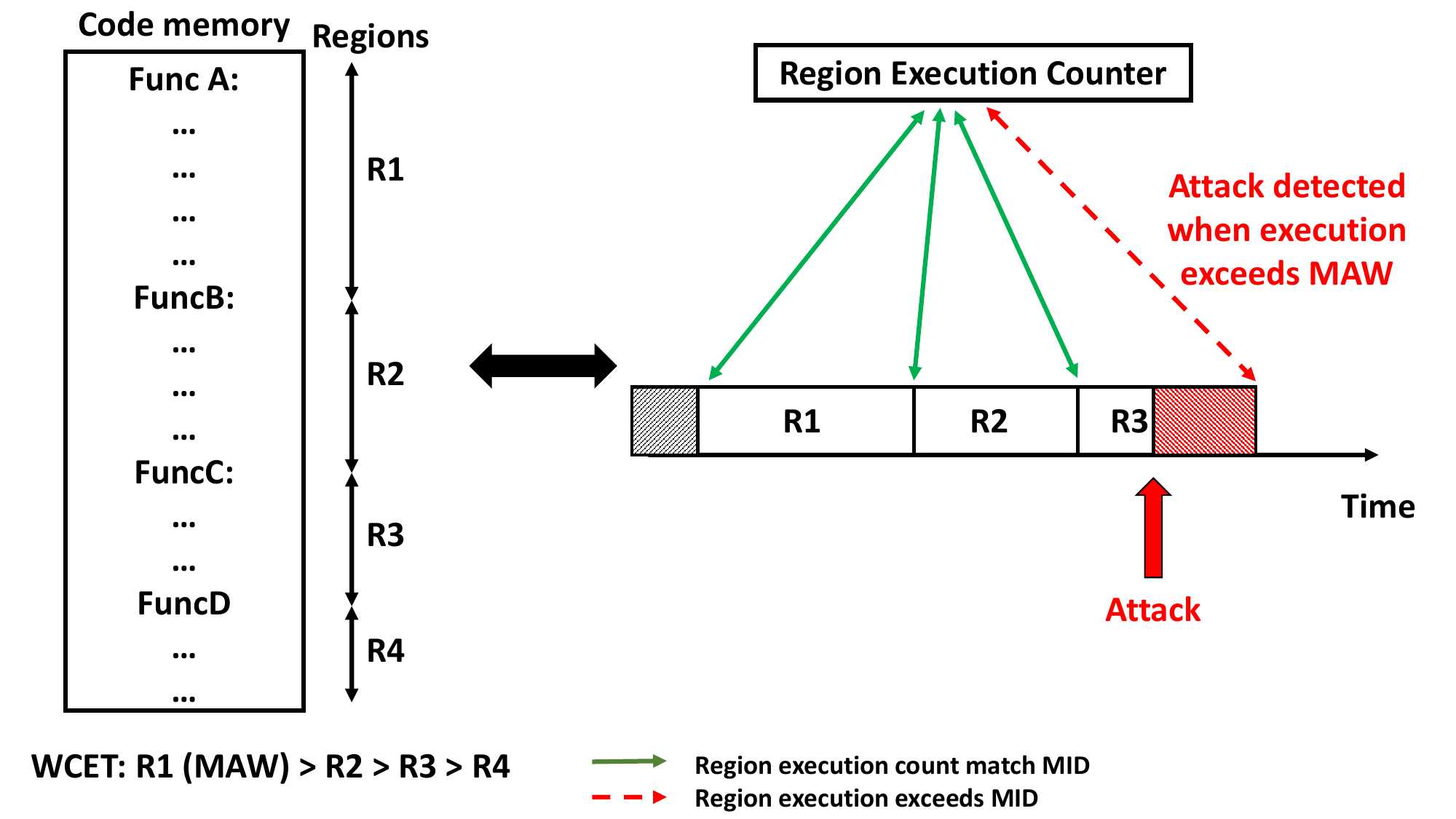}
\caption{Utilizing execution time (MID) as a metric to determine control-flow attacks (Section~\ref{sec:timing_deviations}).}
\label{fig:timingdeviations}
\end{figure}

\subsubsection{Utilizing WCET:} 
\label{sec:bellec}
While RECFISH implements well known CFI techniques, Bellec et.al.'s~\cite{bellec2020attack} proposal utilizes the predictability of real-time systems to detect control-flow violations. An overview of the approach is provided in Figure~\ref{fig:timingdeviations}. Their approach is based on the simple idea that an attacker will cause a control-flow violation to perform some malicious action. This will undoubtedly cause an increase in execution time, over and above the execution time of the system's tasks. Since real-time systems have well-defined task timing parameters, it is within reason to expect that an attacker-controlled execution would show a marked increase in execution time. A monitoring mechanism could, theoretically, detect such an increase and expose an attacker. The authors are able to support such a mechanism by first splitting the code base, consisting of a single task, into \textit{regions}. Regions are either non-overlapping or located entirely within another region. Since the WCET of the task is known, each region within the task code is assigned a WCET of its own, called the maximal inner duration (MID). The MID of a region does not include the MID of a sub-region. Therefore, the sum of MIDs of all regions covering a task's code, would equal the task's WCET. The authors define another metric called the maximal attack window (MAW). For a set of monitored regions, the MAW is the maximum MID of that set. Therefore, the goal is to find the best possible set of regions such that a) the entire task code is covered, and b) the MAW is minimized. The authors perform a search, bounded by the available memory to store region boundaries as well as performance metrics during runtime, to find the best possible set of regions. To evaluate their approach, the authors propose a custom hardware architecture that can detect when code execution enters and exits a region, as well as keep constant track of the time the processor spends within a region. If the time spent exceeds the MAW, an attack would be detected. The authors utilized two benchmark suites, M\"{a}larden, and Polybench. They found that their approach had a mean latency of ~95\% (maximum of ~99\%) of the MAW before it detected an attack, where the MAW sizes ranged from few hundred up to over 160,000 CPU cycles. However, they found that their approach calculated MAWs of 600 or fewer CPU cycles for half of the benchmarks. 

Due to the detection latency, this approach has similar issues as those that utilize laziness, specifically, the attacker could damage the system before it is detected. Further, it requires extensive modifications to the architecture to support it. However, it presents an interesting starting point for CFI mechanisms that effectively utilize the predictability of a real-time system to inform their approach.

\subsubsection{Timing code in hard real-time context}
\label{sec:ecfi}
We end our discussion of the state-of-the-art CFI for real-time systems with Abbasi et.al's~\cite{ecfiabasi} ECFI. ECFI is built for Programmable Logic Controllers (PLC) which are commonly found as the computing units for industrial-control systems. ECFI is a middle-ground approach, utilizing coarse-grained or fine-grained (depending on whether the code has pointer-based calls) CFI as well as exploiting the high predictability of the typical hard real-time system where PLCs serve as computational units, to detect if an attack causes a sudden increase in execution time to warrant the need to perform CFI checks. ECFI operates by capturing control-flow data in a global shadow-stack during system execution, and then check the data in a low-priority process. ECFI presents an amalgamation of traditional CFI techniques and utilization of predictability of the time-domain.

\red{Note that there are related techniques to improve the schedulability of security mechanisms in general, such as Hasan et al.'s Contego framework~\cite{hasan2017contego} that introduces the concept of abstract security tasks into the system, but such techniques are not specifically designed for CFI and are not directly compatible with any of the work presented in this section.} 

\subsection{\red{Section Summary and Observations}}

\red{Our discussion of CFI techniques for real-time embedded systems is summarized in Table~\ref{tab:table_of_contents}. In general, we see a lack of techniques that consider timing constraints. While prior work has explored applying, with varying degrees, timing constraints to improving CFI schedulability there is still clear room for exploring this domain. For example, none of the techniques presented consider overloaded system conditions, or utilize timing to amortize the cost of CFI in such situations. For example, a periodic real-time system has well-defined intervals of slack. By deferring CFI operations to these slack intervals, it would be possible to reduce the effective in-line overhead that the CFI operation introduces while executing the system application, an observation we also state in Section~\ref{sec:resource_constrained_summary}. In our survey of CFI techniques for real-time systems, we have not found any technique that capitalizes on system slack in this manner. On the other hand, Hao et al.'s technique in Section~\ref{sec:exhaustive_search} while useful to reduce the cost of CFI to maintain schedulability, can be considered incomplete in terms of security since only a subset of the code executed at runtime is actually checked. This could be exploited by a smart attacker, especially one aware of the technique used to decide which jobs do not have CFI checks. Some mitigation could be provided by randomizing the schedule using techniques such as using Yoon et al.'s TaskShuffler~\cite{yoon2016taskshuffler}, but even such works have been shown to be defeated by carefully crafting an attack~\cite{nasri2019pitfalls} that defeats the randomization. Essentially we do not see novel techniques that successfully use real-time constraints to amortize the cost of a \textit{complete implementation} of CFI for real-time systems. Bellec et al's approach could be considered as a good starting point for creatively using timing constraints, however, it has its own failings which we discuss in Section~\ref{sec:bellec}.}

\section{Summary and Open Challenges}
\label{sec:summary_challenges}

For convenience, special terminology/mechanisms names that have been discussed before are listed here alongside the relevant section in the paper:

\textbf{Silhouette} - Section~\ref{sec:advanced_shadow}, \textbf{Lazy} - Section~\ref{sec:cfl}, \textbf{Timing deviation} - Section~\ref{sec:timing_deviations}, BBB-CFI - Section~\ref{sec:forwardedge}, \textbf{RECFISH, ECFI} - Section~\ref{sec:cfi_rtos}, \textbf{Context-sensitive} - Section~\ref{sec:forwardedge}

\begin{table}[]
\begin{tabular}{|l|l|}
\hline
\multicolumn{1}{|c|}{Category}                                                                                                                        & \multicolumn{1}{c|}{Technique and Summary}                                                                                                                                                                                                                                                                                             \\ \hline
\begin{tabular}[c]{@{}l@{}}Implementation:\\ Standard CFI techniques on\\ different architectures\end{tabular}                                        & \begin{tabular}[c]{@{}l@{}}1) Silhouette - Shadow stack and binary labeling on ARM Cortex-M\\ 2) RECFISH - Shadow stack and binary labeling on ARM Cortex-R \end{tabular}                                                                                                                  \\ \hline
\begin{tabular}[c]{@{}l@{}}Design Changes:\\ Non-standard CFI techniques\\ utilizing standard control-flow \\ start and end points\end{tabular}       & \begin{tabular}[c]{@{}l@{}}1) Control-Flow Locking - Lazy control-flow evaluation. Single technique \\ for forward and backward-edge\\ 2) uRAI - Collapse shadow stack into a single register using XOR operations.\\ 3) Zipper Stack - Custom hardware to collapse shadow stack into a single register \\ via HMAC operations.\end{tabular} \\ \hline
\begin{tabular}[c]{@{}l@{}}Modern hardware architecture:\\ Techniques that utilize new \\ processor architecture features\end{tabular}                & \begin{tabular}[c]{@{}l@{}}1) CFI Care -  Shadow stack hidden by ARM TrustZone \\ 2) TZmCFI - Nested interrupts (RTOS) aware shadow stack in ARM TrustZone.\\ 3) PACStack - ARM pointer authentication (ARMv8.3-A) utilized for collapsing \\ shadow stack in single register\end{tabular}                                                \\ \hline
\begin{tabular}[c]{@{}l@{}}Underlying Principle:\\ CFI techniques that detect \\ control-flow deviations using\\ non-standard principles\end{tabular} & \begin{tabular}[c]{@{}l@{}}1) Timing deviation - Detect WCET violation of code segments using custom hardware\\ 2) ECFI - Built for PLCs. Detects timing violations code during runtime\end{tabular}                                                                                                                                   \\ \hline
\end{tabular}
\caption{A summary of techniques discussed in depth in Section~\ref{sec:cfi_resource_constrained} and Section~\ref{sec:cfi_real-time}}
\label{tab:summary}
\end{table}
\red{ 
A summary of our discussion in prior sections is presented in Table~\ref{tab:summary}. Some common themes and omissions in the techniques presented are:
\begin{enumerate}
  \item Most prior CFI work utilize some form of software-hardware bypass to accommodate hardware constraints present in resource-constrained embedded systems. The techniques trade-off performance and security to create the best possible compromise for their target hardware architecture. 
  \item The wide variety and heterogeneity of embedded system hardware make it difficult to all but qualitatively compare techniques in terms of memory and performance. Many require the use of custom/bespoke hardware architectures such as Zipper Stack (which requires a custom HMAC and special registers to speed up CFI). It is, therefore, difficult to judge if one technique is better. The applicability of any of the approaches we list for embedded and real-time embedded systems is dependent on the target application. We, therefore, only provide some qualitative discussion and summary, especially for the techniques discussed in-depth for embedded systems in Section~\ref{sec:cfi_resource_constrained} to aid the reader. 
  \item With the exception of Bellec et al.'s work~\cite{bellec2020attack}, the design of \textit{conventional} CFI for embedded and/or real-time embedded systems can primarily be viewed as \textit{memory-based}, where CFI is performed by detecting deviations from expected instruction memory accesses. There is no fundamental difference in the detection methodology across all the presented techniques.
  \item Real-time CFI mechanisms, other than techniques such as that presented by Bellec et al.'s work~\cite{bellec2020attack}, are \textit{ad-hoc} in design. None of the techniques seem to utilize the strict timing requirements of the system to aid CFI. CFI, in essence, is detecting deviation in system behavior and timing critical systems depend heavily on being temporally correct. However, utilizing temporal guarantees exclusively to detect abnormal behavior can reduce the effectiveness of the mechanism as we discuss in Section~\ref{sec:timing_deviations}  In fact, for a real-time embedded system, some assumptions can be made (we will discuss a possible approach later in this section) that can synergistically aid conventional CFI and improve its performance. 
\end{enumerate}
}

In this section we first present open challenges to the real-time community based on our understanding of the state-of-research in CFI for real-time embedded systems. We also present general consideration points that have not yet been incorporated into CFI designs.

\subsection{Real-time challenges}

We believe there exists two broad avenues of research that could be undertaken immediately, considering the state-of-the-arts. 

\noindent \textbf{Bounded laziness:} CFI designs for real-time systems are few in number and do not seem to capitalize on system predictability. In particular, laziness, such as that introduced by control-flow locking is a promising mechanism for hard real-time systems due to its ability to defer CFI checks. However, a drawback of their approach, and Bellec et.al's timing deviation based mechanism (Section~\ref{sec:timing_deviations}) is the lack of expressiveness in the threat model, specifically, the time at which an attacker is able to affect the system. For example, the proposed mechanisms fail to consider that an attacker could modify and produce system outputs, such as sending messages via a network controller to other systems, before the CFI mechanism detects an attack.
\red{On the other hand, conventional CFI techniques have an unnecessary sense of \textit{urgency} since CFI is performed as close to when control-flow path changes as possible. For example, the mechanism presented in Silhouette adapts well-known CFI techniques which all perform CFI during a control-flow transfer event. We believe there is a middle-ground that can improve performance and still maintain the \textit{usefulness} of CFI. That is, the purpose of CFI to detect an attacker before they are able to damage the system, is still maintained. This is because real-time systems inherently have discrete and well-known time instances where they must generate system output. For example, a typical control-system in an industrial environment, would command an actuator after a defined interval of time. In such instances, there is no need to \textit{urgently} perform CFI, but CFI work can be deferred to much later by recording the control-flow transfer event and then verifying at a later stage before the actuator command is sent out. The advantage of such mechanisms would be a reduction in temporal overload situation which plagues current CFI implementation since they must be performed while system code is executing. However, the tradeoff is increased memory usage to record control-flow events. We believe there is ample opportunity to capture such memory-timing-security tradeoff situations in real-time systems due to the higher degree of predictability over general systems.} Essentially, there is a need to define how lazy CFI can be, and develop system/task models that enforce these boundaries.

\noindent \textbf{Multi-thread/core scheduling:} RECFISH and ECFI showcases the applicability of well-known CFI techniques to multi-threaded hard real-time systems. We believe there is an opportunity to extend the concept of bounded laziness to multi-threaded systems and utilize the large pool of available real-time scheduling theory to tighten the bounds. For example, there could be an opportunity to steal system slack for performing CFI operations. In the case of multi-core scheduling, cores could be dedicated to performing CFI operations. From a security perspective, ECFI implicitly trusts the scheduler's integrity. However, in advanced threat models where an attacker could have the privilege to disrupt scheduler operations, such as modifying the system timer to warp the scheduler's sense of time, such defense mechanisms could fail. Prior work to secure time sources, such as TimeSeal~\cite{anwar2019securing} could provide some inspiration to solve this problem. 

\noindent \textbf{Determining CFI related workload attributes: } Addressing the previous challenges would also require determining the real-time properties of CFI operations, such as the WCET of CFI operations, or how CFI operations would be incorporated into advanced real-time models such as those that consider varying task periods, varying number of real-time tasks, the effect of servicing aperiodic tasks, etc. The WCET of CFI too could be difficult to accurately determine especially if the mechanism operates on historical control-flow data, such as in context-sensitive CFI, where the amount of data can vary during system operation.

\subsection{General challenges}

In addition to the real-time system specific challenges listed above, there are some general considerations that should be incorporated into future designs. \red{The following challenges are not just limited to CFI mechanisms but the system security research in general.}

\noindent \textbf{Power consumption:} An often overlooked component of embedded system development is power consumption. This is also evident in every CFI design reviewed in this paper. None of the mechanisms consider power consumption, which is especially important in embedded systems operating off batteries and deployed in the field. \red{Some designs such as that provided by Das et al.~\cite{das2016fine} provide power consumption measurements of their custom control-flow checking hardware design implemented on an FPGA. However, such measurements are an exception rather that the norm with respect to CFI research. Custom designs presented by other work such as Zipper Stack~\cite{li2020zipper} do not provide information regarding power consumption, making it difficult to decide applicability of such work to severely power-constrained and hard real-time environments such as heart pacemakers.} Since CFI techniques such as shadow stacks have high memory access rates, impacts on system power consumption of different techniques must be considered. 

We believe that, alongside real-time scheduling theory, techniques such as Dynamic Voltage Frequency Scaling (DVFS), \red{backed by an extensive pool of scheduling algorithms that utilize DVFS~\cite{saha2012experimental,safari2018pl}, could provide significant reduction in system power consumption and interesting schedulability issues. Interestingly, a logical correlation can be made between coarse-grained CFI and reduced power consumption, by virtue of reduction of CFI checks that are required due to the coarseness of the design. A study on the relation between coarse CFI and power consumption of design on commercial-off-the-shelf hardware could have an immediate impact within the research community, providing researchers guidance on which type of CFI and what aspects of CFI design have the worst effects on power consumption. We also believe that a new class of schedulability-power co-design problems could arise from utilizing laziness in CFI to limit the peak power consumption of a system by carefully differing CFI to low power consumption phases of the system.}

The goal of CFI is similar to that of system reliability improvement techniques, i.e., to prevent incorrect execution and/or detect when incorrect execution occurs. There is a large amount of prior work that discuss mechanisms to implement recovery schemes with minimal impact to system power consumption. Such work could be used as inspiration to create energy-aware CFI mechanisms.

\noindent \textbf{Portability:} In general, CFI for resource-constrained embedded systems adapt well-known CFI techniques such as shadow stacks and labeling to such systems while working around their limitations. A primary observation is that many of these workarounds are very specific to the hardware platform that the authors target. For example, Silhouette targets ARMv7-M and therefore modifies store instructions to use the MPU on this architecture. Since these limitations are hardware-specific, designing realistic CFI mechanisms for such systems that are also portable is difficult. Unlike desktop or server-grade hardware, where commodity systems usually include processors with similar underlying architecture, embedded system utilize architectures from ARM, RISC, MIPS, etc. as well as application-specific designs. Designing a one-size-fits-all mechanism for such a wide-range of target architectures is a difficult challenge. Further, architectures such as ARM are very modular, allowing hardware vendors a high-degree of flexibility to add or remove features to adjust manufacturing costs and provide a wide portfolio of devices at every price point. There is, thus, a need to design feasible CFI mechanisms that operate completely in software (or with minimal hardware requirements), to allow for portable designs. However, the overhead of such designs remains to be seen.

\noindent \textbf{Advanced CFI and beyond:} As discussed in Section~\ref{sec:forwardedge}, there is a need to consider context-sensitivity in real-time embedded systems to thwart attacks that can bypass even fine-grained CFI. We are not aware of the existence of such techniques. Finally, there is a gap in research for embedded and real-time embedded systems regarding state-of-the-art data oriented programming~\cite{hu2016data} (DOP) attacks. These do not redirect control-flow, but attack program data, such as the counter variable used for a loop. Such attacks cannot be mitigated using any of the CFI designs discussed in this paper since they do not cause deviations in control-flow path. \red{Note that techniques such as timing deviation detection discussed in Section~\ref{sec:timing_deviations} may be able to detect such attacks, but the assumption here is that the attacker is knowledgeable and does not violate the MAW during an attack.} Data oriented attacks are powerful and have been shown to be capable of influencing program output as well as disclose private information.

\section{Conclusion}

We have examined multiple CFI schemes in the paper, starting from the core mechanisms that help enforce CFI, to the necessary workarounds required to support them in resource-constrained embedded environments. We have also looked at the modifications necessary to support real-time schedulers and how real-time characteristics can be effectively utilized for CFI. While CFI has been adopted by higher-end systems, designs for resource-constrained embedded systems are still mostly academic and not yet widely deployed due to unmanageable performance overhead in some cases. As we have seen CFI will undoubtedly have overhead due to hardware constraints, but techniques such as laziness that trade-off detection speed with overhead could provide an interesting avenue for future work.



\bibliographystyle{ACM-Reference-Format}
\bibliography{ref}


\end{document}